\documentclass[12pt]{article}
\usepackage[dvips]{graphicx}
\begin{document}
% \newcommand{\eproof}{\rule{0.2cm}{0.2cm}}
%%\noindent
%%%%%%%% DEFINITIONS by Yu LUCHKO
%\newcommand{\FTS}[2]{\frac{{\textstyle #1}}{{\textstyle #2}}}
%\newcommand{\NN}{I\!\!N}
%\newcommand{\ka}{I\!\!K}
%\newcommand{\rgr}{{\rm grad}}
%\newcommand{\CC}{I\!\!\!\!C}
%\newcommand{\RR}{I\!\!R}
\newcommand{\intl}{\int\limits}
\newcommand{\suml}{\sum\limits}
\setcounter{page}{1}
\tolerance=10000
\def\theequation{\arabic{section}.\arabic{equation}}
\font\title=cmbx12 scaled\magstep2 \font\bfs=cmbx12 scaled\magstep1
\font\sc=cmcsc10
 %%%%%%%%%%%%%%%%%%%%% DEFINITIONS
 %\def\pni{\par \noindent}
\def\pni{\par\noindent}
\def\vsh{\smallskip}
\def\vs{\medskip}
\def\vvs{\bigskip}
\def\vvvs{\bigskip\medskip} %% {\vskip 1.5truecm}
\def\vsp{\vsh\pni} %% ie. \smallskip + \par
\def\vsn{\vsh\pni}
\def\cen{\centerline}
\def\ra{\item{a)\ }} \def\rb{\item{b)\ }}   \def\rc{\item{c)\ }}
\def\q{\quad} \def\qq{\qquad}
\def \rec#1{{1\over{#1}}}
\def\ds{\displaystyle}
\def\eg{{\it e.g.}\ }
\def\ie{{\it i.e.}\ }
\def\versus{{\it vs.}\ }
%%%%%%%%%%%%%%%%%%% DEFINIZIONI SIMBOLI MATEMATICI %%%%%%%%%%%%%%%%%%%%
\def\e{{\rm e}}
\def\d{\partial}
\def\dx{\partial x}    \def\dt{\partial t}
\def\Ai{{\rm Ai}\,}
\def\Erfc{{\rm Erfc}\,}
\def\u{\widetilde{u}}
\def\ul{\widetilde{u}} %%% Laplace Transform  (LT)
\def\uf{\widehat{u}} %%% Fourier Transform  (FT)
\def\r{\right} \def\l{\left}
\def\rt{\right} \def\lt{\left}
\def\lra{\Longleftrightarrow}
\def\RR{\vbox {\hbox to 8.9pt {I\hskip-2.1pt R\hfil}}}
\def\NN{{\rm I\hskip-2pt N}}
\def\CC{{\rm C\hskip-4.8pt \vrule height 6pt width 12000sp\hskip 5pt}}
\def\L{{\mathcal L}} %%% Laplace Transform !!!!
\def\Gz{\Gamma(z)}    \def\Ga{\Gamma(\alpha)}
\def\Gaz{\Gamma(\alpha\,,\, z)}
\def\gaz{\gamma(\alpha\,,\, z)}
\def\DG{{{D}}_\Gamma}
 \def\E{{\rm E}}
\def\Ai{{\rm Ai}\,}
\def\Erfc{{\rm Erfc}\,}
\def\Ei{{\rm Ei}\,}
\def\Ein{{\rm Ein}\,}
\def\log{{\rm log}\,}
\def\EE{{\mathcal E}}
\def\Gz{\Gamma(z)}    \def\Ga{\Gamma(\alpha)}
\def\Gaz{\Gamma(\alpha\,,\, z)}
\def\gaz{\gamma(\alpha\,,\, z)}
\def\DG{{{D}}_\Gamma}
 \def\E{{\rm E}}
\def\Ai{{\rm Ai}\,}
\def\Erfc{{\rm Erfc}\,}
\def\Ei{{\rm Ei}\,}
\def\Ein{{\rm Ein}\,}
\def\log{{\rm log}\,}
\def\EE{{\mathcal E}}
%%%%%%%%% for LAPLACE - FOURIER TRANSFORMS %%%%%%%%
 \def\bar{\widetilde}
\def\hatt{\widehat}
\def\epsilons{{\widetilde \epsilon(s)}}
\def\sigmas{{\widetilde \sigma (s)}}
\def\fs{{\widetilde f(s)}}
\def\Js{{\widetilde J(s)}}
\def\Gs{{\widetilde G(s)}}
\def\Fs{{\widetilde F(s)}}
 \def\Ls{{\widetilde L(s)}}
 \def\Hs{{\widetilde H(s)}}
 \def\Ks{{\widetilde K(s)}}
\def\L{{\mathcal L}} %%% Laplace Transform !!!!
\def\F{{\mathcal F}} %%% Fourier Transform !!!!
\def\M{{\mathcal M}} %%% Mellin Transform !!!!
\def\P{{\mathcal{P}}} %%% Probability !!!!
\def\H{{\mathcal{H}}} %%% FOX Kernel !!!!
%%%%%%%%% SETS of NATURAL, REAL, COMPLEX NUMBERS : \NN, \RR, \CC
\def\NN{{\bf N}}
\def\RR{{\bf R}}
\def\CC{{\bf C}}
\def\ZZ{{\bf Z}} %%%%%%%% !!!!!!!!!!!!
%%%%%%%%%%%%% INTEGRAL and DIFFERENTIAL OPERATORS
\def\I{{\cal I}}  %% INTEGRAL
\def\D{{\cal D}}  %% DERIVATIVE
%%%%
\def\erf{\hbox{erf}}  \def\erfc{\hbox{erfc}}
%%% displayed fractions %%%%%%%%%%%%%%%%%%%%%%%
\def\dfrac#1#2{\displaystyle{\frac {#1}{#2}}}
%%% definition of displayed fractions; example:
% $\dfrac{a^3}{\Gamma(\dfrac{\lambda}2)} $
%%%%%%%%%%%%%%%%%%%%%%%%%%%%%%%%%%%%%%%%%%%%%%%%%%%%%%%%
%%%%%%% making running heads %%%%%%%%%%%%%%%%%%%%%%%%%%%
% \markboth
% {\rm \centerline {F. Mainardi, R. Gorenflo}}
% {\rm \centerline{Time-fractional derivatives in relaxation processes}}
%%%%%%% making running heads on even and odd pages %%%%%%%
\cen{{\bf FRACALMO PRE-PRINT: \    www.fracalmo.org}}
\vsh
\cen{{\bf Fractional Calculus and Applied Analysis,
  Vol. 10 No 3 (2007) 269-308}}
\vsh
\cen{An International Journal for Theory and Applications \ ISSN 1311-0454}
\vsh
\cen{{\bf www.diogenes.bg/fcaa/}}
\vs
\hrule
% \end{center}
%%%%%%%%%%%%%%%%%%%%%%%%%%%%%%%%%%%%%%%%%%%%%%%%%%%%%%%%%%%%%%%%%%%%%%%%%
%%BEGINNING OF TEXT
%%%%%%%%%%%%%%%%%%%%%%%%%%%%%%%%%%%%%%%%%%%%%%%%%%%%%%%%%%%%%%%%%%%%%%%%%
\vskip 0.50truecm
\font\title=cmbx12 scaled\magstep2
\font\bfs=cmbx12 scaled\magstep1
\font\little=cmr10
\begin{center}
{\bfs Time-fractional derivatives in relaxation processes:}
\vs

{\bfs a tutorial survey}

\vvs
{Francesco MAINARDI} $^{(1)}$ and
{Rudolf GORENFLO}$^{(2)}$

%%%%%%%%%%%%%%%%

\vs

$\null^{(1)}$
 {\little Department of Physics, University of Bologna, and INFN,} \\
              %%%  Sezione di Bologna, \\
{\little Via Irnerio 46, I-40126 Bologna, Italy} \\
%% Tel: +39-051.2091098 $\;$ Fax: +39-051.247244 $\;$\\
{\little Corresponding Author. E-mail: {\tt francesco.mainardi@unibo.it}} 
\\ [0.25 truecm]
$\null^{(2)}$
 {\little Department of Mathematics and Informatics, Free University  Berlin,} \\
{\little Arnimallee 3, D-14195 Berlin, Germany} \\
%%%%%%%%%%%%%%%% Phone: +49 30 4504 2247 \\
{\little E-mail: {\tt gorenflo@mi.fu-berlin.de}}
\\ [0.25 truecm]
%% $^{2}$ Fachbereich Mathematik \& Informatik \\
%% Erstes Mathematisches Institut\\
%% Freie Universitaet Berlin, Arnimallee 3, \\
%% D-14195 Berlin, GERMANY \\[6pt]
%% e-mail: gorenflo@mi.fu-berlin.d
%% $^{2}$ Department of Mathematics and Informatics \\
%% Free University of Berlin\\
%% Arnimallee 3, D-14195 Berlin, GERMANY \\[6pt]
%% e-mail: gorenflo@mi.fu-berlin.de

\end{center}
%%%%%%%%%%%%%%%%%%%%%%%%%%%%%%%%%%%%%%%%%%%%%%%%%%%%%%%%%%

%% \fnote {} {$^*$ This paper is dedicated to
%% Rudolf Gorenflo,  Professor Emeritus
%% of Mathematics at
%% the Free University of Berlin,
%% on the occasion of his 70-th birthday.}
\vsp
\centerline{{\sc Dedicated to Professor Michele Caputo, Accademico dei Lincei, Rome,}}
\\
\centerline{{\sc on the occasion of his 80-th birthday (May 5, 2007).}}

\vskip 0.5truecm

\begin{abstract}
\noindent
The aim of this tutorial survey is to revisit the basic theory of
relaxation processes governed by linear differential equations of
fractional order. The fractional derivatives are  intended both in
the Rieamann-Liouville sense and in the Caputo sense. After giving a
necessary outline of the classical theory of linear viscoelasticity,
we contrast these two types of fractional derivatives in their
ability to take into account  initial conditions in the constitutive
equations of fractional order. We also provide historical notes on
the origins of the Caputo derivative and on the use of fractional
calculus in viscoelasticity.
%\end{abstract}
\end{abstract}
\vsp
{\it 2000 Mathematics Subject Classification}:
26A33,  %%%%  (main);    Fractional derivatives and integrals
33E12, %% Mittag-Leffler type functions
33C60,  %% hypergeometric integrals and functions defined by them
44A10,  %% Laplace Transforms
45K05,  %% integro-partial differential equations
%% 74Dxx, %%  Materials of strain-rate type and history type, other materials
%% with memory (including elastic materials with viscous damping, various viscoelastic materials)
74D05, %%  Linear constitutive equations
%% 74D10 Nonlinear constitutive equations
%% 74D99 None of the above, but in this section
\vsp

%%%%%%%%%%%%%%%%%%%%%%%%%%%%%%%%%%%%%%%%%%%%%%%%%%%%%

%\section*{Introduction}
\section*{Introduction}

 In recent decades the field of fractional
calculus has attracted  interest of researchers in several areas
including mathematics, physics, chemistry, engineering  and  even
finance and social sciences. In this  survey
 we revisit the fundamentals of fractional calculus
 in the framework of the most simple time-dependent processes
 like those concerning relaxation phenomena.
  We devote particular attention to the technique of Laplace
transforms  for treating the operators of differentiation
of non integer order
(the term "fractional" is kept only for historical reasons),
  nowadays known as Riemann-Liouville
and Caputo derivatives.  We shall
point out the fundamental role of
the Mittag-Leffler function (the Queen function of the fractional calculus),
whose main properties are reported in an {\it ad hoc} Appendix.
The topics discussed here will be:
(i) essentials  of fractional calculus
  with basic formulas for Laplace transforms (Section 1);
(ii) relaxation  type differential equations
    of fractional order (Section 2);
(iii) constitutive equations of fractional order in viscoelasticity (Section 3).
The last topic is treated in detail  since, as a matter of fact,
the linear theory of viscoelasticity is the field
where we find the most extensive applications of {fractional calculus}
already since a long time, even if often only in an implicit way.
Finally, we devote Section 4 to historical notes concerning
the origins of the Caputo derivative and the use of fractional calculus
in viscoelasticity in the past century.

 %%%%%%%%%%%  Section 1 %%%%%%%%%%%%%%%%%%%%%%%%%%%%%%%%%
 % \setcounter{section}{1}\setcounter{equation}{0}

\section{Definitions and properties}

For a sufficiently well-behaved function $f(t)$ (with $ t\in \RR^+$)
we may define the  derivative
of a positive non-integer order %% $\mu  $ ($m-1 <\mu \le m\,,$ $\, m\in \NN$),
in two different senses,  that we refer here as to
{\it Riemann-Liouville} (R-L) derivative
and {\it Caputo} (C) derivative, respectively.
\vsp
Both derivatives are related to the so-called Riemann-Liouville
fractional integral. For  any $\alpha >0$ this fractional integral is  defined as
$$ J_t^\alpha  \, f(t) :=    \rec{\Gamma(\alpha )}\,
\int_0^t\!  (t-\tau)^{\alpha-1} \, f(\tau )\, d\tau\,,
\eqno(1.1)  $$
where ${\ds \Gamma(\alpha):= \int_0^\infty \e^{-u}\, u^{\alpha-1}\, du}$ denotes the Gamma function.
For existence of the integral (1) it is sufficient that the
   function $f(t)$ is locally integrable in $\RR^+$ and for
   $t \to 0$ behaves like $O(t^{-\nu})$ with a number $\nu < \alpha$.
%%%%%%%%%%%%%%%%%%%
\vsp
For completion we define  $J_t^0 = I$ (Identity operator).
\vsp
We recall the semigroup property
$$ J_t^\alpha \, J_t^\beta = \,
   J_t^\beta  \, J_t^\alpha = J_t^{\alpha +\beta} \,, \quad
 \alpha ,\beta \ge 0\,. \eqno(1.2)$$
Furthermore we note that for $\alpha \ge 0$
$$ J_t^{\alpha  }\, t^{\gamma}=
   {\Gamma(\gamma +1)\over\Gamma(\gamma +1+ \alpha  )}\,
     t^{\gamma+\alpha  }\,,
  \q \gamma >-1\,. %%% \q t>0\,.
\eqno (1.3)
$$
The fractional derivative of order $\mu >0$
in the {\it Riemann-Liouville} sense  is defined as the operator
$D_t^\mu$ which is the
left inverse of
the Riemann-Liouville integral of order $\mu $
(in analogy with the ordinary derivative), that is
$$ D_t^\mu \, J_t^\mu  = I\,, \q \mu >0\,. \eqno(1.4) $$
If $m$ denotes the positive integer
such that  $m-1 <\mu  \le m\,,$
  we recognize from Eqs. (1.2) and (1.4):
$$D_t^\mu \,f(t) :=  D_t^m\, J_t^{m-\mu}  \,f(t)
\,, \quad m-1 <\mu \le m \,.\eqno (1.5)$$
In fact, using the semigroup property (1.2), we have
$$ D_t^\mu \, J_t^\mu = D_t^m\, J_t^{m-\mu}\,  J_t^\mu = D_t^m\, J_t^m = I\,.$$
Thus (1.5) implies
$$\,
 \!\! \! \! D_t^\mu  \,f(t) =
 \cases{
{\ds {d^m\over dt^m}}\lt[
  {\ds
  \rec{\Gamma(m-\mu )}\int_0^t
    \! {f(\tau)\,d\tau  \over (t-\tau )^{\mu  +1-m}}
    }\rt],  & $m-1<\mu<m;$ \cr\cr
   {\ds {d^m \over dt^m} f(t)}, &$\mu=m .$}
    \eqno(1.5')
   $$
For completion we define $ D_t^0 = I\,. $
\vsp
On the other hand, the fractional derivative of order $\mu $ in the
{\it Caputo} sense  is defined as the operator
$\,_*D_t^\mu$  such that
$$    _*D_t^\mu \,f(t) :=  \, J_t^{m-\mu } \, D_t^m \,f(t)
\,, \quad m-1 <\mu \le m \,.
\eqno(1.6)$$
This implies
$$\,
    _*D_t^\mu\,f(t) =
    \cases{
    {\ds \rec{\Gamma(m-\mu )}}\,{\ds\int_0^t
 \! {\ds {f^{(m)}(\tau)\, d\tau \over (t-\tau )^{\mu  +1-m}}}} \,,& $m-1<\mu<m\,;$\cr\cr
   {\ds {d^m \over dt^m} f(t)} \,, & $\mu=m\,.$}
  \eqno(1.6')
   $$
Thus, when the order is not integer the two fractional derivatives
 differ in that  the standard derivative of order $m$
does not generally commute with the fractional  integral. Of course
the Caputo derivative (1.6$'$) needs higher regularity conditions of
$f(t)$ than the Riemann-Liouville derivative (1.5$'$). \vsp
%%%%%%
We point out that the  {\it Caputo} fractional derivative
  satisfies the  relevant property
of being zero when applied to a constant, and, in general,
to any power function  of non-negative integer degree less than $m\,,$
if its order $\mu $ is such that $m-1<\mu \le m\,. $
Furthermore we note for $\mu \ge 0$:
$$ D_t^{\mu }\, t^{\gamma}=
   {\Gamma(\gamma +1)\over\Gamma(\gamma +1-\mu )}\,
     t^{\gamma-\mu }\,,
  \q \gamma >-1\,. \eqno(1.7)
$$
It is instructive to compare Eqs. (1.3), (1.7).
 \vsp
In \cite{GorMai CISM97}
we have shown the essential  relationships between the two fractional
derivatives %% of the same  non-integer order
%% when both of them exist
for the same  non-integer order %%% $\mu \in (m-1,m)$,
$$  _*D_t^\mu  \,f(t)   =  \cases{
  {\ds \, D_t^\mu  \lt[ f(t) -
  \sum_{k=0}^{m-1} f^{(k)}(0^+)\,{t^k\over k!} \rt]} , \cr\cr
 {\ds  D_t^\mu  \, f(t) -
    \sum_{k=0}^{m-1} {f^{(k)}(0^+) \,
t^{k-\mu }\over \Gamma(k-\mu+1)} },\cr } \q m-1<\mu<m\,.
\eqno(1.8)$$ In particular we have from (1.6$'$) and (1.8)
$$
\begin{array}{ll}
& \,_*D_t^\mu   f(t)  =
{\ds \rec{\Gamma(1-\mu )}\,\int_0^t
 \! {f^{(1)}(\tau) \over (t-\tau )^{\mu}}\, d\tau} \\
  &= {\ds \, D_t^\mu \,\lt[ f(t) - f(0^+) \rt]}  =
{\ds \, D_t^\mu\,f(t) - f(0^+)\, {t^{-\mu} \over\Gamma(1-\mu)}}\,,
\end{array}
 \; 0<\mu <1\,.
  \eqno(1.9)$$
%% $$  _tD_*^\mu  \,f(t) = \,
%%   {\ds _tD^\mu \,\lt[ f(t) -   f(0^+) - f^{(1)}(0^+) t\rt] } =
% {\ds _tD^\mu\,f(t) - {f(0^+)\, t^{-\mu} \over\Gamma(1-\mu)} -
% {f^{(1)}(0^+)\, t^{-1-\mu} \over\Gamma(2-\mu)}  }\,, \,
% 1<\mu <2\,. \eqno(9b)$$
%%%%%%%%%%%%%%%%%%%%%
The {\it Caputo} fractional derivative represents a sort of
regularization in the time origin for the {\it Riemann-Liouville}
fractional derivative. We note that for its existence all the
limiting   values $f^{(k)}(0^+):= {\ds \lim_{t\to 0^+} D_t^{k}f(t)}$
are required to be finite for $k=0,1,2,\dots, m-1$. In the special
case   $f^{(k)}(0^+)=0$  for $k=0,1,  m-1$, the two fractional
derivatives coincide.
%% The {\it Caputo} fractional derivative is of course more restrictive
%% than the {\it Riemann-Liouville} fractional derivative
%% in that   the derivative of order $m$  is required to exist
%% and subjected to some regularity conditions.
\vsp We observe the different behaviour  of the two fractional
derivatives at the end points of the interval $(m-1,m)\,$ namely
when the order is any positive integer, as it can be noted from
their definitions (1.5), (1.6). In fact, whereas for $\mu \to m^-$
both derivatives reduce to $D_t^m$, as stated in Eqs. (1.5$'$),
(1.6$'$),
 due  to the fact that the operator $J_t^0 \,=\, I$ commutes with $D_t^m$,
 for $\mu \to (m-1)^+$ we have \vskip -8pt
$$
 \!\! \mu \to (m-1)^+
 :
 \cases{
 {\ds D_t^{\mu} f(t)} \to
 {\ds \,\, D_t^m \, J_t^1 \, f(t) =  D_t^{(m-1)} \,f(t)= f^{(m-1}(t)},
 \cr \cr
{\ds \,_*D_t^{\mu} f(t)} \to
{\ds J_t^1\, D_t^m \, f(t) =  f^{(m-1)}(t) - f^{(m-1)}(0^+) }.
 }
  \eqno(1.10)
$$
As a consequence, roughly speaking, we can say that $D_t^{\mu}$ is, with respect to its order $\mu\,, $
 an operator continuous  at any positive integer, %%  (by definition),
whereas $\, _*D_t^{\mu}$   is an operator only left-continuous.
\vsp
The above  behaviours have induced us to keep for the Riemann-Liouville
derivative the same symbolic notation as for the standard derivative of integer order,
while for the Caputo derivative to decorate the corresponding symbol with  subscript $*$.
%% The last limit can be formally obtained
%% by recalling the formal representation of the $m$-th derivative of the
%% Dirac  function,
%% $\delta^{(m)} (t) = t^{-m-1}/\Gamma(-m)\,,$ $\, t \ge 0\,,$
%% see  \cite{Gel'fand 64}. %% \eg Gel'fand and Shilov
\vsp
% \medskip
We also note, with  $m-1 < \mu \le m\,,$ and $c_j$ arbitrary
constants,
$$   \hskip -0.5truecm  D_t^\mu  \, f(t) \,=\, D_t^\mu   \, g(t)
   \,  \Longleftrightarrow  \,
  f(t) = g(t) + \sum_{j=1}^m c_j\, t^{\mu -j} \,,  \eqno(1.11)
    $$ \vspace*{-10pt}
$$    _*D_t ^\mu  \, f(t) \,=\,  _*D_t^\mu   \, g(t)
   \,  \Longleftrightarrow  \,
  f(t) = g(t) +  \sum_{j=1}^m c_j\, t^{m-j} \,. \eqno(1.12)
     $$
 Furthermore, we observe that in case of a non-integer order for both fractional derivatives
 the semigroup property (of the standard derivative for integer
 order) does not hold for both fractional derivatives when the order is not integer.
\vsp
%% Last but not least,
We point out the major utility
of the Caputo fractional derivative
in treating initial-value problems for physical and engineering
applications where initial conditions are usually expressed in terms of
integer-order derivatives. This can be easily seen
using the Laplace transformation\footnote{%
The Laplace transform of a well-behaved function $f(t)$
 is defined as
 $$\widetilde f(s) =
\L \lt\{ f(t);s\rt\}
 := {\ds \int_0^{\infty}} \! \e^{\ds \, -st}\, f(t)\, dt\,, \quad
 \Re\,(s) > a_f\,. $$ %% s \in \CC\,.$$
 We recall that under suitable conditions the Laplace transform of the
 $m$-derivative of $f(t)$ is given by
 $$ \L \lt\{ D_t^m \,f(t) ;s\rt\} =
      s^m \,  \widetilde f(s)
   -\sum_{k=0}^{m-1}    s^{m  -1-k}\, f^{(k)}(0^+) \,,
\quad f^{(k)}(0^+) :=  \lim_{t\to 0^+} D_t^{k}f(t)\,.$$
}
% \vfil \ % newpage
%%%
\vsp
For the Caputo derivative of order $\mu$ with $ m-1<\mu  \le m $
we have
$$
\begin{array}{ll}
& {\ds \L \lt\{ _*D_t^\mu \,f(t) ;s\rt\}}  =
     {\ds s^\mu \,  \widetilde f(s) -\sum_{k=0}^{m-1}    s^{\mu  -1-k}\, f^{(k)}(0^+)} \,,\\ \\
 & f^{(k)}(0^+) := {\ds \lim_{t\to 0^+} D_t^{k}f(t)\,.}
 \end{array}
\eqno(1.13)$$
%% \vsp
The corresponding rule for the Riemann-Liouville
derivative of order $\mu$   is
%% for $m-1<\mu \le m $ it reads
$$
\begin{array}{ll}
 & {\ds \L \lt\{ D_t^\mu  \, f(t);s\rt\}} =
     {\ds s^\mu \,  \widetilde f(s) -\sum_{k=0}^{m-1} s^{m -1-k}\, g^{(k)}(0^+)}\,, \\ \\
&  g^{(k)}(0^+) := {\ds \lim_{t\to 0^+} D_t^{k}g(t)\,,} \quad
\hbox{where}\quad  g(t):= J_t^{(m-\mu)}\,f(t) \,.
\end{array}
\eqno(1.14)$$
 Thus it is more cumbersome to use the rule  (1.14)
than (1.13). The rule (1.14) requires initial values  concerning
an extra function $g(t)$ related to the given $f(t)$
through a  fractional integral.
%%   \q m-1<\mu  \le m \,, \eqno(14)$$
%% In analogy with (13),  the limit for $t \to 0^+$
%% is understood to be taken after the operations of fractional integration
%% and derivation.
%% \vsp
However, when all the limiting   values $f^{(k)}(0^+)$
for $k=0,1,2,\dots$
are finite and the order is not integer, we can prove
 that  the corresponding $g^{(k)}(0^+)$ vanish so that
 %%%  (as noted by  Mainardi \cite{Mainardi }FDA04}),
 formula (1.14)  simplifies into
$$ \L \lt\{ D_t^\mu  \, f(t);s\rt\} =
      s^\mu \,  \widetilde f(s) \,,
      \quad  m-1 <\mu< m\,.
      \eqno(1.15)$$
      For this proof it is sufficient to
      apply the Laplace transform to the second equation (8),
      by recalling that
      $ \L \lt\{ t^\alpha ;s\rt\} =
     \Gamma(\alpha+1)/ s^{\alpha+1} \,$ for $\alpha >-1$,
     and then to compare  (1.13) with (1.14).
%%%%%%%%%%%%%%%%
\vsp
For fractional differentiation on the positive semi-axis
we  recall another %% , namely aider
 definition for the fractional derivative
recently introduced by Hilfer,
see \cite{Hilfer 00}  %%  2000; p. 113
and  \cite{Seybold-Hilfer FCAA05},
which interpolates the previous definitions (1.5) and (1.6).
Like the two derivatives previously discussed, it  is related to
a  Riemann-Liouville integral.
  In our notation
it reads
$$ D_t^{\mu,\nu} := J_t^{\nu(1-\mu)}\, D_t^1 \,J_t^{(1-\nu)(1-\mu)} \,,
\q 0<\mu<1\,,\q 0\le \nu \le 1\,.  \eqno(1.16)$$
We can refer it to as the {\it Hilfer} (H) fractional derivative of order $\mu$ and type $\nu$.
The Riemann-Liouville derivative corresponds to the type   $\nu =0$
whereas that Caputo derivative to the type $\nu =1$.
%%%%%%%%%%
\vsp
We have here not discussed the Beyer-Kempfle approach investigated
and used in several papers by Beyer and Kempfle et al.:
this approach is appropriate for causal processes not starting at a finite instant of time,
see e.g.
\cite{Beyer-Kempfle ZAMM95,Kempfle-Schaefer-Beyer NLD02}.
%% (Beyer and Kempfle, 1995), Kempfle and Sch\"afer, 2002).
They define the time-fractional derivative on the whole real line as a pseudo-differential operator
via its Fourier symbol.
%% We will give details on this concept in an extension of this memorandum.
The interested reader is referred to the above mentioned papers and references therein.
 \vsp
 For further reading %% more details
on the theory and applications of fractional integrals and derivatives
(more generally of fractional calculus)
we may recommend e.g. our CISM Lecture Notes
\cite{Gorenflo CISM97,GorMai CISM97,Mainardi CISM97},
%% Gorenflo and Mainardi,  1997; Gorenflo, 1997; Mainardi 1997)
the review papers
\cite{Metzler-Klafter PhysRep00,Metzler-Klafter 02,Sokolov-Klafter-Blumen FK02},
and  the books
%% (Oldham \& Spanier, 1974), (Samko, Kilbas \& Marichev, 1993), (Miller \& Ross, 1993),
%%   (Kiryakova, 1994), (Podlubny, 1999),(Hilfer, 2000),  (West, Bologna \& Grigolini, 2003),
%% (Zaslavsky, 2005), (Magin, 2006), (Kilbas, Srivastava \& Trujillo, 2006).
\cite{Miller-Ross BOOK93,Kiryakova BOOK94,Hilfer BOOK00,Kilbas-et-al BOOK06,%%
Magin BOOK06,Oldham-Spanier BOOK74,SKM 93,West BOOK03,Zaslavsky BOOK05}
with references therein.

  %%%%%%%%%%% Section 2  %%%%%%%%%%%%%%%%%%
 % \setcounter{section}{2}\setcounter{equation}{0}
 \section{Relaxation equations of fractional order}
 %%%%%%%%%%%%%%%%%%%%%%%%%%%%%%%%%%%%%%%%%%

The different roles played by the R-L and C derivatives and by the intermediate H derivative
 are  clear  when one wants to consider the corresponding fractional generalization
 %% of to the initial value problem for
 of  the first-order differential equation governing the
 phenomenon of (exponential) relaxation.  Recalling   (in non-dimensional units)
 the initial value problem %% standard equation
 $$\frac{du}{dt } = -u(t) \,, \quad t\ge 0\,, \quad \hbox{with}\quad u(0^+)= 1\,\eqno(2.1)$$
 whose solution is
 %% Th well-known solution of the above initial value problem
 $$u(t) = \exp (-t)\,,\eqno (2.2)$$
  the following  three alternatives with respect to the R-L and C fractional derivatives
  with $\mu \in (0,1)$ are offered in the literature:
   $$ _*D_t^\mu \,u(t) = -u(t)\,, \quad t\ge 0\,, \quad \hbox{with}\quad u(0^+)= 1\,.\eqno(2.3a)$$
  $$ D_t^\mu \,u(t) = -u(t) \,,\quad t\ge 0\,, \quad \hbox{with}\quad \lim_{t\to 0^+} J_t^{1-\mu}\,u(t)= 1\,,\eqno(2.3b)$$
   $$\frac{du}{dt } = - D_t^{1-\mu}\,u(t) \,,\quad  t\ge 0\,, \quad \hbox{with}\quad u(0^+)= 1\,,\eqno(2.3c)$$
 %%% EX Fig 1
 \vsp
 In analogy to the standard problem (2.1) we  solve these three problems
  with the Laplace transform technique,  using respectively the rules (1.13), (1.14) and (1.15).
 The problems (a) and (c) are equivalent since  the Laplace transform of the solution
 in both cases comes out as
 $$  \widetilde u(s) = \frac{s^{\mu-1}}{s^\mu +1}\,, \eqno(2.4) $$
 whereas in the case (b) we get
 $$  \widetilde u(s) = \frac {1}{s^\mu +1}=
  1- s \frac{s^{\mu-1}}{s^\mu +1}  \,.\eqno(2.5)$$
 The Laplace transforms in (2.4)-(2.5) can be expressed in terms of  functions
 of Mittag-Leffler type, for which we provide essential information in Appendix.
 In fact, in virtue of the Laplace transform pairs,
 that here we report from Eqs. (A.4) and (A.8),
%%   see e.g. (Podlubny, 1999),
$$ {\mathcal{L}}\{ E_\mu (-\lambda t^\mu);s\} =
\frac{s^{\mu -1} } {s^\mu + \lambda}\,,  \eqno(2.6)$$
$$ \L\{t^{\nu-1}\, E_{\mu,\nu} (-\lambda t^\mu);s\} =
\frac{s^{\mu -\nu} }{s^\mu + \lambda}\,,  \eqno(2.7)$$
with $\mu, \nu \in \RR^+$ and $\lambda \in \RR$,
we have: in the cases (a) and (c),
$$ u(t) = \Psi(t) := E_\mu(-t^\mu)\,, \quad t\ge 0\,, \quad 0<\mu<1\,,\eqno(2.8)$$
and in the case (b), using the identity (A.10),
 $$ \!\!u(t)= \Phi(t) := t^{-(1-\mu )}
  E_{\mu ,\mu} \left(-  t^{\mu }\right)
=  -  \frac{d}{dt} E_\mu  \left (- t^{\mu }\right),
\; t\ge 0,
\;  0<\mu \le 1.\eqno(2.9)$$
It is evident that for $\mu \to 1^-$ the solutions of the three initial value problems
(2.3) reduce to the standard exponential function (2.8).
\vsp
We  note that the case (b) is of little interest from a physical
view point since the corresponding  solution (2.9) is infinite in the time-origin.
%%%%%%%%%%%%%%%
%%%%%%%%%% FIGURE 1
\begin{figure}[h!]
\begin{center}
 \includegraphics[width=.75\textwidth]{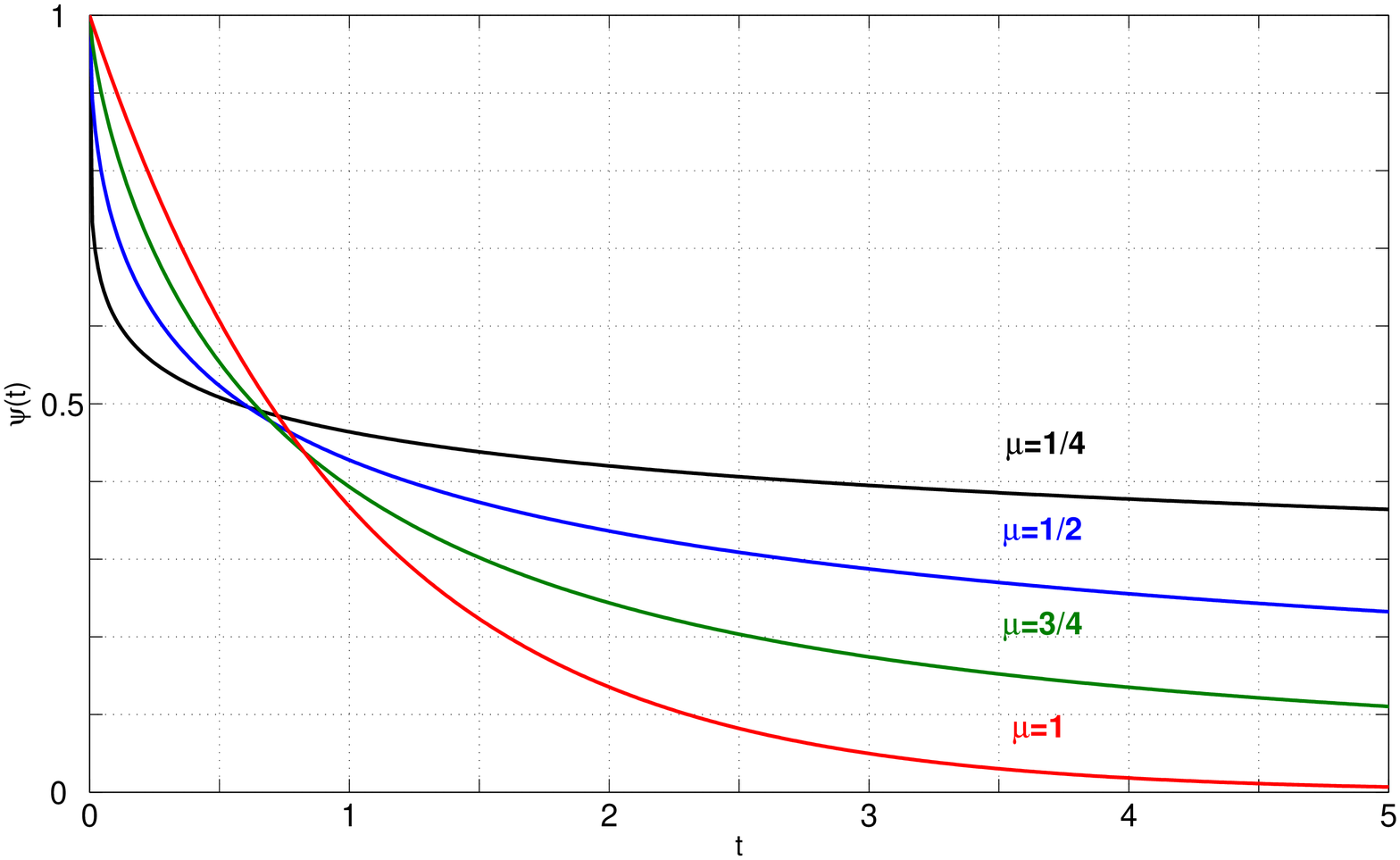}
 \includegraphics[width=.75\textwidth]{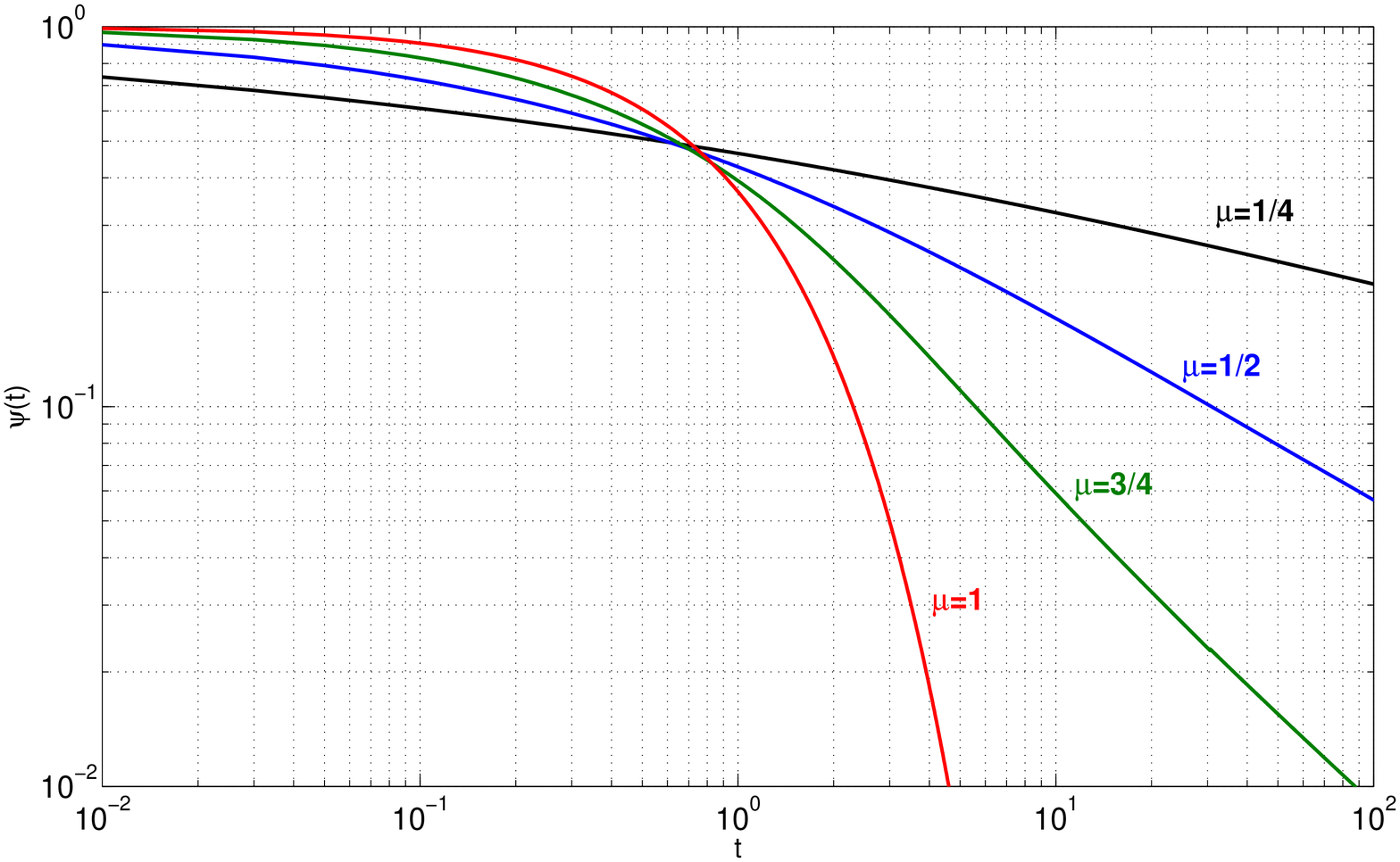}
\end{center}
\vskip -1.0truecm
 \caption{Plots  of the %% Mittag-Leffler type
 function  $\Psi(t)$  with $\mu=1/4, 1/2, 3/4, 1$ versus $t$;
 top: linear scales ($0\le t \le 5$);
 bottom: logarithmic scales ($10^{-2} \le t \le 10^{2}$).
 }
 \end{figure}
%%%%%%%%%  END OF FIGURE 1
 \vsp
We recall that former plots of the Mittag-Leffler function $\Psi(t)$
are found (presumably for the first time in the literature) in the
1971 papers by Caputo and Mainardi \cite{Caputo-Mainardi
PAGEOPH71}\footnote{Ed. Note: This paper has been now reprinted in
this same FCAA issue, under the kind permission of  Birkh\"{a}user
Verlag AG.} and \cite{Caputo-Mainardi RNC71} in the framework of
fractional relaxation for viscoelastic media,
in times when such function was almost  unknown. %%  function
Recent numerical treatments of the Mittag-Leffler functions have
been provided by Gorenflo, Loutschko and Luchko \cite{GoLoLu 02}
with {\it MATHEMATICA}, and by Podlubny \cite{Podlubny MATLAB06}
with {\it MATLAB}.
%%%%%%%%%%%%
%%%%%%% FIGURE 2 %%%%%%%%%%%%%%
\vsp
\begin{figure}[h!]
\begin{center}
 \includegraphics[width=.75\textwidth]{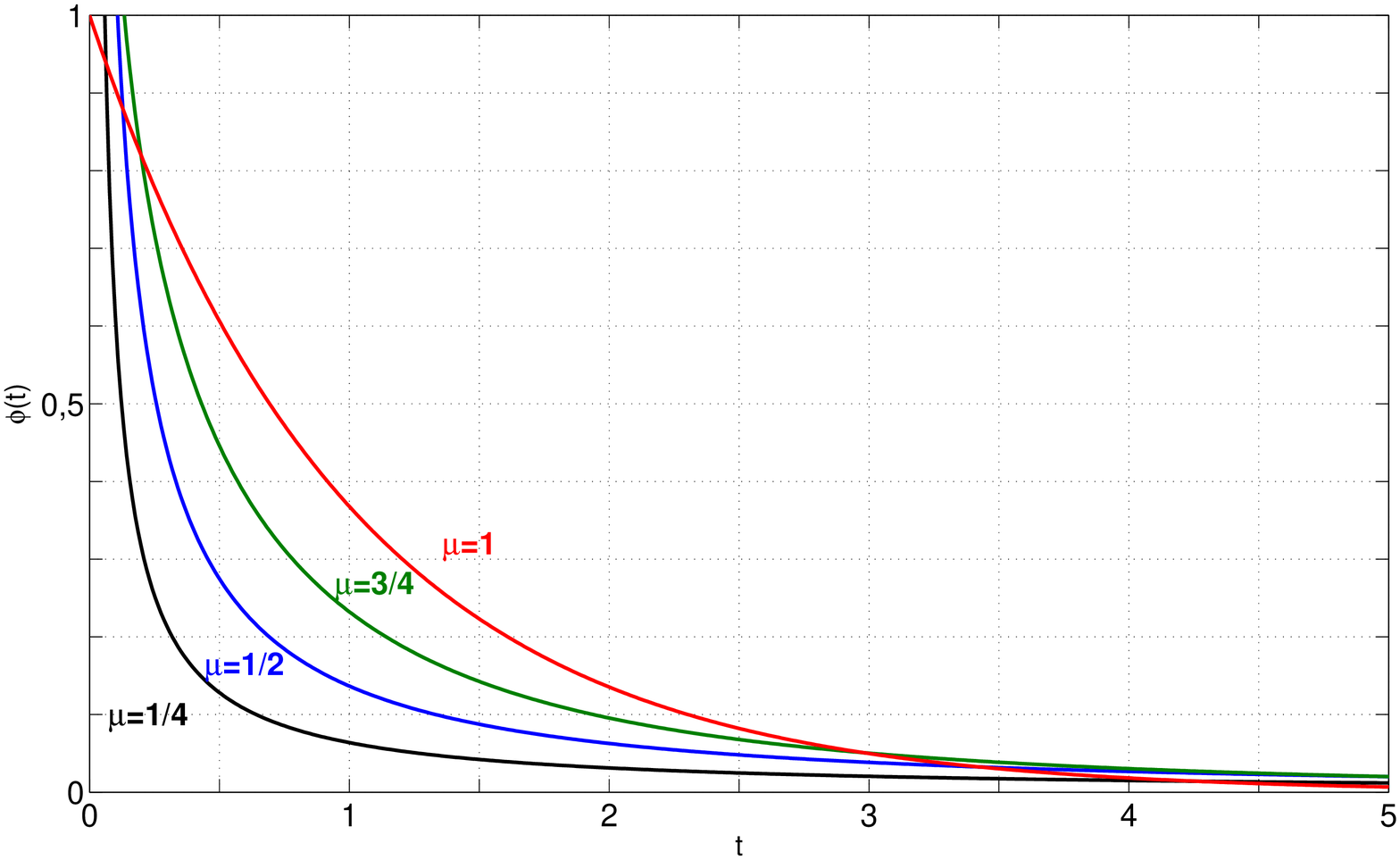}
 \includegraphics[width=.75\textwidth]{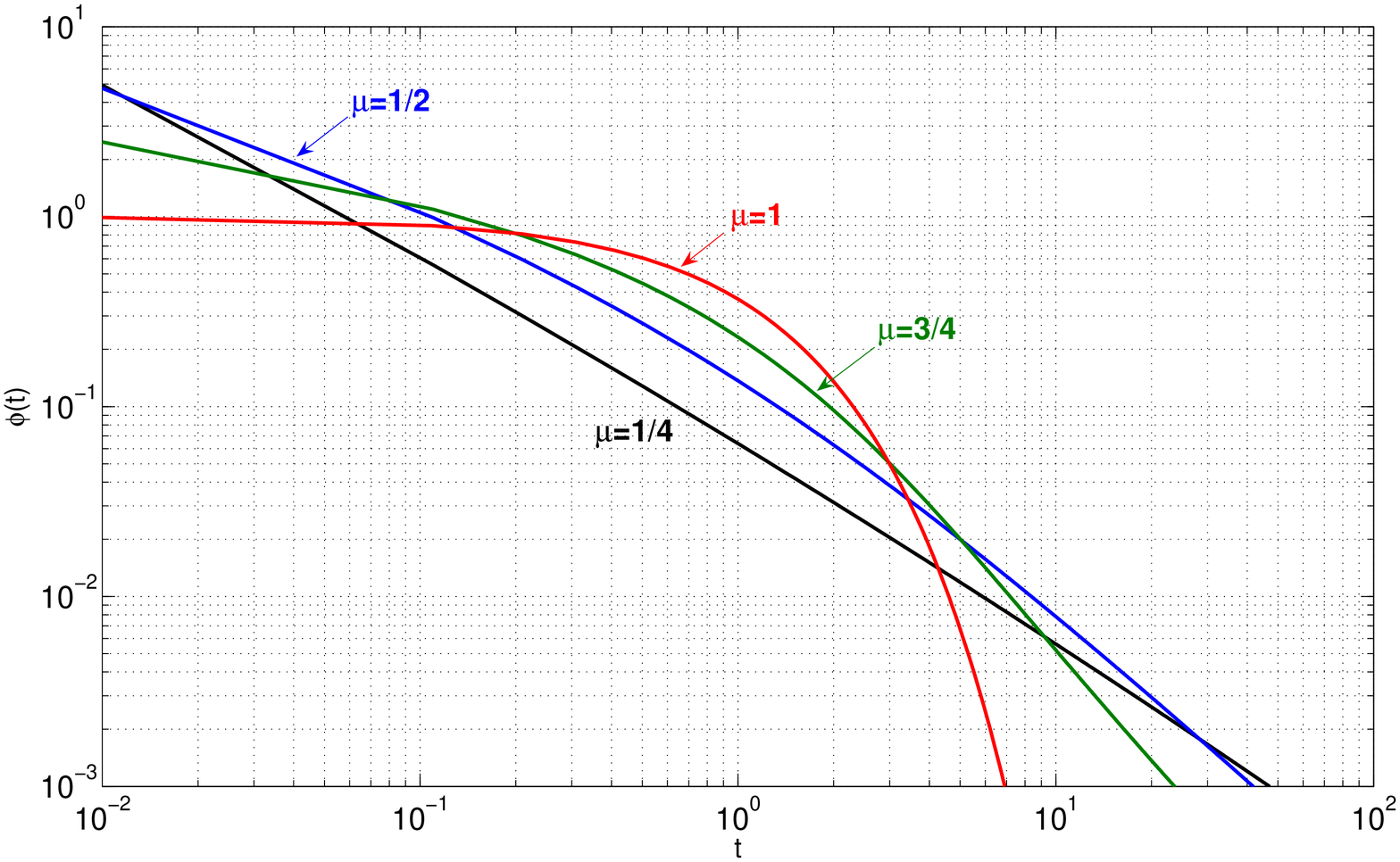}
\end{center}
 \vskip -0.8truecm
 \caption{Plots  of the %% Mittag-Leffler type
 function $\Phi(t)$  with $\mu= 1/4, 1/2, 3/4, 1$ versus $t$;
 top: linear scales ($0\le t \le 5$);
 bottom: logarithmic scales ($10^{-2} \le t \le 10^{2}$).
 }
 \end{figure}
 %%%%%%%%%%% END OF FIGURE 2
\vsp
The plots of the functions $\Psi(t)$ and  $\Phi(t)$ are shown in
Figs. 1 and 2, respectively, for some rational values of the
parameter $\mu$, by adopting linear and logarithmic scales.
\vsp 
For the use of the Hilfer intermediate  derivative in
fractional relaxation we refer to Hilfer himself, see \cite{Hilfer
00}, p.115, according to whom the Mittag-Leffler type function
$$ u(t)= t^{(1-\nu)(\mu-1)}\, E_{\mu, \mu+\nu(1-\mu)}(-t^\mu)\,, \quad t\ge 0\,,\eqno(2.10)$$
is the solution of %% the fractional relaxation equation
 $$ D_t^{\mu,\nu} \,u(t) = -u(t) \,,\quad t\ge 0\,, \quad \hbox{with}
 \quad \lim_{t\to 0^+} \,J_t^{(1-\mu)(1-\nu)}\,u(t)= 1\,.\eqno(2.11)$$
 In fact, Hilfer has shown that the Laplace transform of the solution of (2.11)
 is   %%  (in our notation)
 $$\widetilde u(s) = \frac{s^{\nu(\mu-1)}}{s^\mu + 1} \,,\eqno(2.12)$$
 so,  as a consequence of (2.7), we find (2.10).
%%%%%%%%%
For plots of the function in (2.10) we refer to \cite{Seybold-Hilfer FCAA05}.
%% In particular our previous results are included in his analysis
%%  if we change sign in front of the RHS of (19a), (19b),
%% the solution (24) can be seen as the eigen-function of the C derivative
%% and     the solution (25) as the eigen-function of the RL derivative.
%%%%%%%%%%%%%%%%% \newpage
%\vspace*{-5pt}
 %%%%%%%%%%%% Section 3 %%%%%%%%%%%%%%%%%%%%%%%%%%%%%%%%%%%%%%%%%%

% \setcounter{section}{3}\setcounter{equation}{0}
 \section{Constitutive equations of fractional order in  viscoelasticity}
 %%%%%%%%%%%%%%%%%%%%%%%%%%%%%%%%%%%%%%%%%%%%%%%%%%%%%%%%%%%%%%
% \vspace*{-5pt}

 In this section  we present the fundamentals of  linear
 Viscoelasticity restricting our attention to the one-axial case
and assuming that the viscoelastic body is quiescent for all times prior to
some starting instant that we assume as $t=0$.
\vsp
For the sake of convenience both stress $\sigma(t)$ and strain $\epsilon(t)$ are intended
to be normalized, \ie scaled  with respect to a suitable reference state
$\{\sigma _0\,, \,\epsilon _0\}\,. $
After this necessary introduction we shall consider the main topic
concerning viscoelastic models based on differential constitutive equations of fractional order.

%%%%%%%%%%%%%
%\subsection{Generalities} %%% 3.1.
% \newpage %%% \vspace*{-8pt}
\subsection{Generalities} % \vspace*{-6pt}
%%%%%%%%%%%%%

According to the linear theory, the viscoelastic body
can be considered as a linear system
with the stress (or strain) as the excitation function (input)
and the strain (or stress) as the response function (output).
 In this respect, the response functions to
an excitation expressed by the Heaviside step function $\Theta(t)$
are known  to   play a fundamental role both from a mathematical
and physical point of view. We denote by $J(t)$ the strain response
to the unit step of  stress,  according to the {\it creep test}
 and by $G(t)$ the stress response to a unit step of strain,
according to the {\it relaxation test}.

\vsp
The functions $J(t)\,,\, G(t)$  are usually referred to as the
{\it creep compliance} and {\it relaxation modulus}
respectively, or, simply, the {\it material functions}
of the viscoelastic body.   In view of the causality
requirement, both  functions are  causal, \ie  vanishing
for $t<0$.
%%%%%%%%
\vsp
The limiting values of the material functions
for $t \to 0^+$ and $t \to +\infty$  are related to the
instantaneous (or glass) and equilibrium behaviours of the viscoelastic
body, respectively. As a consequence, it is usual to denote
   $J_g := J(0^+)$  the {\it glass compliance},
   $ J_e := J(+\infty)$ the {\it equilibrium  compliance},
and
  $ G_g := G(0^+)$ the {\it glass modulus}
  $ G_e := G(+\infty)$ the {\it equilibrium modulus}.
As a matter of fact, both the material functions are
non-negative. Furthermore, for  $0< t < +\infty\,,$
 $ J(t)$ is a  {\it non decreasing function} and
 $G(t)$ is a  {\it non increasing function}.
%%%%%%%%%%
\vsp
The  monotonicity properties of $J(t)$ and $G(t)$
are  related respectively   to the physical phenomena
of  strain {\it creep} and stress {\it relaxation}\footnote{%%
For the  %% more restrictive
mathematical conditions that  the material functions must
satisfy  to agree with the most common experimental observations
(physical realizability) we refer to the recent paper
by Hanyga  \cite{Hanyga_05c} and references therein. We also note that in some cases
the material functions can contain terms  represented by {\it generalized functions}
(distributions) in the sense of Gel'fand-Shilov \cite{Gelfand-Shilov_64}
or {\it pseudo-functions} in the sense of Doetsch \cite{Doetsch_74}}.
%% \vsp
Under the hypotheses of causal histories, we get the stress-strain
relationships
$$
\cases{
{\ds \epsilon (t)}
 = {\ds \int_{0^-}^t  \!\! J(t-\tau)\, d\sigma (\tau )}
 = {\ds \sigma (0^+)\, J(t) + \int_0^t  \!\! J(t-\tau)\,
  \frac{d}{d\tau }\sigma (\tau ) \, d\tau }\,,\cr
{\ds \sigma  (t)}
= {\ds \int _{0^-}^t  \!\!G(t-\tau )\, d\epsilon  (\tau )}
=  {\ds \epsilon(0^+)\, G(t) + \int_0^t  \!\! G(t-\tau)\,
 \frac{d}{d\tau } \epsilon  (\tau ) \, d\tau}\,,}
 \eqno(3.1)$$
 where the passage to the RHS is justified
 if differentiability is assumed for the stress-strain histories,
 see also the excellent book by Pipkin \cite{Pipkin BOOK86}.
%% Unless and until we find it makes any sense to do otherwise,
%% we implicitly restrict our attention to causal histories.
Being of convolution type,  equations (3.1) can be conveniently
treated by the technique of Laplace transforms so they read
in the Laplace domain
$$ \widetilde \epsilon (s) = s\, \widetilde{J}(s) \, \widetilde \sigma(s)\,,
\quad
\widetilde \sigma  (s) = s\, \widetilde{G}(s) \, \widetilde \epsilon (s)\,,
\eqno(3.2)$$
from which we derive the {\it reciprocity relation}
  $$ s\, \widetilde{J}(s)  = \frac{1}{s\,\widetilde{G}(s)}
 \,. \eqno(3.3)$$
Because of the limiting theorems for the Laplace transform, we deduce that
$ J_g = {1}/ {G_g}$, $\, J_e = {1}/{G_e}$,
with the convention that $0$ and $+\infty$ are reciprocal to each other.
The above remarkable relations allow us to classify the viscoelastic bodies
according to their instantaneous and equilibrium responses
in four types as  stated by Caputo \& Mainardi in their 1971 review paper
 \cite{Caputo-Mainardi RNC71} in Table 3.1.
%%%%        TABLE 3.1 %%%%%%%%%%%%
 \vskip 0.25truecm
\begin{center}
\begin{tabular}{|c||c|c||c|c|}
\hline
% {} &{} &{} &{} &{} \\[-1.5ex]
Type & $J_g$ & $J_e$ & $G_g$ & $G_e$ \\
% \colrule
\hline
I   & $>0$ & $<\infty$ & $<\infty$ & $>0$ \\
II  & $>0$ & $=\infty$ & $<\infty$ & $=0$ \\
III & $=0$ & $<\infty$ & $=\infty$ & $>0$ \\
IV  & $=0$ & $=\infty$ & $=\infty$ & $=0$ \\
\hline
\end{tabular}  %% \label{t2.1}
\vskip 0.25truecm
 Table 3.1 \ The four types of viscoelasticity.
\end{center}
\vskip 0.25truecm
%%%%%%     the end of Table 2.1 %%%%%%

%%%%%%%%%%%%\
\newpage %%%%%%%%%%
\noindent 
We note that the viscoelastic bodies of type I exhibit
both instantaneous and equilibrium elasticity, so their behaviour
appears  close to the purely elastic one for sufficiently short and
long times. The  bodies of  type II and IV exhibit a complete stress
relaxation (at constant strain) since $G_e =0$ and an infinite
strain creep (at constant stress) since $J_e = \infty\,,$ so  they
do not present equilibrium elasticity. Finally, the bodies of type
III and IV do not present instantaneous elasticity  since $J_g =
0\,$ ($G_g =\infty$). Other properties will be pointed out later on.
%% \vsp

%% %%%%%%%%%%
%\subsection{The mechanical models}%% 3.2.
% \vspace*{-8pt} 
\subsection{The mechanical models} %% \vspace*{-6pt}
%%%%%%%%%%%

 To get some feeling for linear
viscoelastic behaviour, it is useful to consider the simpler
behaviour of analog {\it mechanical models}. They are
 constructed from linear springs and dashpots,
disposed singly and in branches of two (in series or in parallel).
As analog of stress and strain, we use the total extending force
and the total extension.
We note that when two  elements are combined in series [in parallel],
their compliances [moduli] are additive. This can be stated
as  a combination rule: {\it creep compliances add in  series,
while relaxation moduli add in parallel}.
%%%%%%%
\vsp
 The mechanical models play an important role in the literature which is
 justified by the historical development.
 In fact, the early theories were established with the aid of these
 models, which are still helpful to visualize properties and laws of the
 general theory, using the combination rule.
Now, it is worthwhile to consider the simple models of Fig. 3
  providing  their governing  stress-strain
relations along with the related material functions.
\vsp
The spring (Fig. 3a) is the elastic (or storage) element, as for it the
force is proportional to the  extension;
it represents a perfect elastic body obeying the Hooke
law (ideal solid).
This model is thus referred to as  the {\it Hooke} model.
If we denote by $m$ the pertinent elastic modulus  we have
$$
 Hooke \; model\;: \q \sigma(t)  = m\, \epsilon (t)\,, \quad\hbox{and}\quad
\cases{
 J(t) = 1/m \,,\cr
 G(t) = m  \,.}
 \eqno(3.4)$$
In this case we have no creep and no relaxation
so the creep compliance and the relaxation modulus are constant functions:
$J(t) \equiv J_g \equiv J_e =1/m$; $G(t)\equiv G_g \equiv G_e =1/m$.
 %%%%%%%%
\vsp
The dashpot (Fig. 3b) is the viscous (or dissipative) element,
the force being proportional to rate of extension;
it represents a perfectly viscous body obeying the Newton law (perfect
liquid).
This model is thus referred to as  the {\it Newton} model.
If we denote by $b$ the pertinent  viscosity coefficient, we have
    $$  Newton \; model\;: \q
\sigma(t)  = b_1\, \frac{d\epsilon}{dt}\,, \quad\hbox{and}\quad
  \cases{
 J(t) = {\ds \frac{t}{b_1}}\,,\cr
 G(t) = b_1 \, \delta (t)\,.
}
 \eqno(3.5)$$
 In this case we have a linear creep $J(t)= J_+t$ and instantaneous relaxation
 $G(t)= G_-\, \delta(t)$ with $G_-= 1/J_+ = b_1$.
 \vsp
We note that
 the {\it Hooke} and {\it Newton} models represent the limiting cases of
viscoelastic  bodies of type $I$ and $IV$, respectively.
\vsp
\begin{figure}[h!]
\begin{center}
 \includegraphics[width=.82\textwidth]{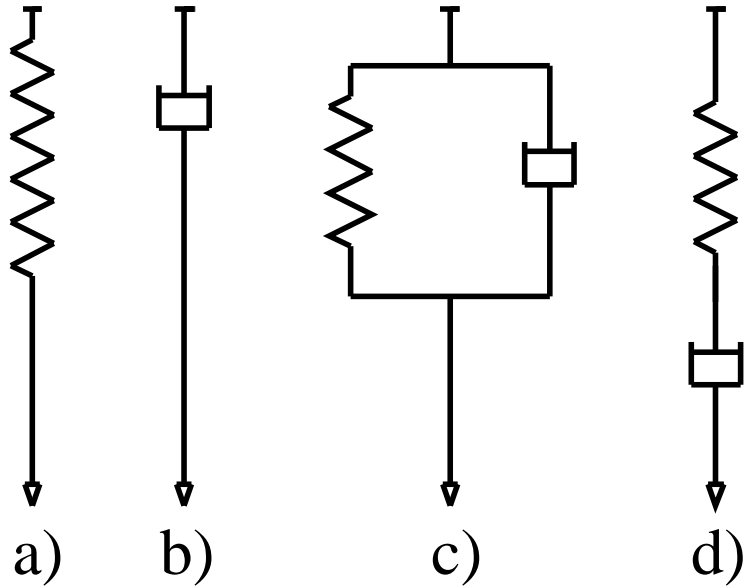}
\end{center}
\caption{The representations of the basic mechanical models:
 a)  spring for Hooke, b) dashpot for Newton,
 c)  spring and dashpot in parallel for Voigt,
 d) spring and dashpot in series for Maxwell.}
\end{figure}
\vsp
A branch constituted by a spring in parallel with a dashpot is known as
  the   {\it Voigt}  model (Fig. 3c). We have
$$ Voigt \; model \,:\;
\sigma(t)  = m\, \epsilon (t) +b_1\, \frac{d\epsilon}{dt}\,, \eqno (3.6)
$$
and
$$\cases{
 {\ds J(t) = J_1 \left[ 1-\e^{\ds - t/\tau _\epsilon}\right]}\,, &
  ${\ds J_1 = \frac{1}{m}\,,\; \tau _\epsilon  = \frac{b_1}{m}}\,,$\cr\cr
 {\ds G(t) = G_e +  G_- \,  \delta(t)} \,,&
   ${\ds G_e=m\,,\; G_- =b_1}\,,$}
 $$
where $\tau _\epsilon$ is referred to as the {\it  retardation time}.
%% $ \alpha \, G_\infty =1\,, \q \tau _\sigma =G_1/G_\infty \,. $
\vsp
A branch constituted by a spring in series with a dashpot is known as
 the  {\it  Maxwell} model  (Fig. 3d). We have
$$
 Maxwell \; model\,: \;
\sigma(t) +a_1 \, \frac{d\sigma}{dt}    = b_1\, \frac{d\epsilon}{dt}\,,
\eqno(3.7)$$
and
$$\cases{
 {\ds J(t) =  J_g + J_+\,t}\,, & ${\ds J_g= \frac{a_1}{b_1} \,,\; J_+ = \frac{1}{b_1}}\,, $ \cr\cr
 {\ds G(t) = G_1\,\e^{\ds -t/\tau_\sigma}}\,, & ${\ds G_1=\frac{b_1}{a_1}\,,\; \tau _\sigma= a_1}\,,$
}$$
where $\tau _\sigma$ is is referred to as  the {\it the relaxation time}.
%% with $ \beta \, J_0 =1\,, \q \tau_\epsilon  =J_0/J_1 \,.$
\vsp
 The {\it Voigt} and the {\it Maxwell} models are thus the simplest
viscoelastic bodies of type   $III$ and $II$, respectively.
The {\it Voigt} model exhibits  an exponential  (reversible)
strain creep but no stress relaxation; it is also referred to
as the retardation element.
The {\it Maxwell} model exhibits  an exponential (reversible) stress
relaxation and a linear (non reversible) strain creep; it is
also referred to as the relaxation element.
\vsp
Based on the combination rule,
 we can continue the previous procedure in order to construct
 the simplest models of type $I$ and $IV$ that require three parameters.
\vsp
 The simplest viscoelastic body of type $I$
 %% requires three parameters, \ie $m$ and  $a\,,\, b$; it
 is obtained by
adding a spring either in series to a Voigt model  or
in parallel to a Maxwell model (Fig. 4a and Fig 4b, respectively).
So doing, according to the combination
rule, we add a positive constant both to the  Voigt-like creep compliance
and to the  Maxwell-like relaxation modulus so that we obtain $J_g >0$
and $G_e >0\,. $
Such a model was
introduced by Zener \cite{Zener BOOK48} with the denomination of {\it Standard
Linear Solid} ($S.L.S.$). We have
$$
Zener \; model \;:\q
\left[1 +a_1 \, \frac{d}{dt}\right] \sigma(t) =
 \left [ m+ b_1\, \frac{d}{dt}\right] \epsilon (t) \,,
 \eqno(3.8) $$
and
$$
\!\!\!
 \cases{
 {\ds J(t) =  J_g + J_1  \left[ 1-\e^{\ds - t/\tau_\epsilon}\right]},
      &
${\ds J_g =  \frac{a_1}{b_1}, \; J_1 =\frac{1}{m}- \frac{a_1}{b_1},\; \tau_\epsilon =\frac{b_1}{m}},$
 \cr\cr
 {\ds G(t) = G_e + G_1 \,\e^{\ds -t/\tau_\sigma} },
        &
 ${\ds G_e =  m, \;
   G_1 = \frac{b_1}{a_1}- m , \;  \tau_\sigma = a_1}\,.$
}$$
%%\vsp
We point out the condition
$ 0< m<b_1/ a_1$
in order  $J_1 ,G_1 $ be positive and hence
$0< J_g <  J_e < \infty $ and $0 <G_e <G_g < \infty\,.$
As a consequence,
we note  that, for the {\it S.L.S.} model,
 the retardation time must be greater than
the relaxation time, \ie
$\, 0 < \tau _\sigma <\tau_\epsilon < \infty \,.$
\vsp
\begin{figure}[h!]
\begin{center}
 \includegraphics[width=.82\textwidth]{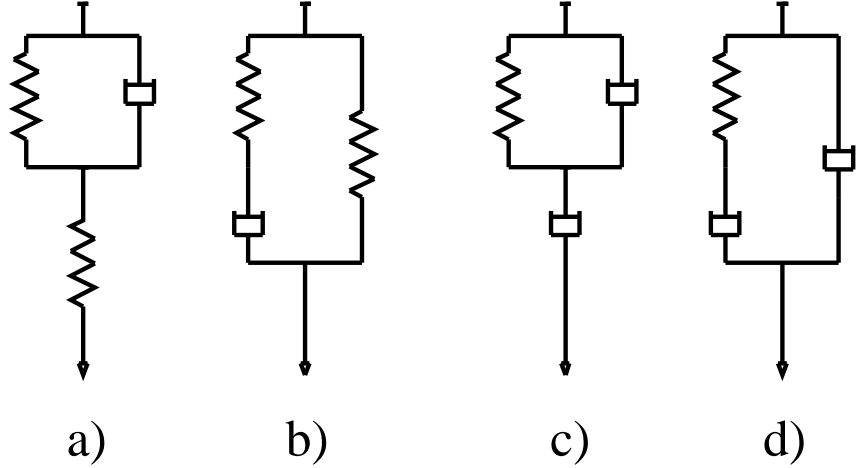}
\end{center}
\caption{The mechanical representations of the Zener [a), b)] and
anti-Zener[c), d)] models:
a)  spring in series with  Voigt, b) spring  in parallel with Maxwell,
c)  dashpot in series with  Voigt, d) dashpot in parallel with Maxwell.}
\end{figure}
\vsp
Also the simplest viscoelastic body of type $IV$ requires three parameters,
\ie   $a_1\,,\, b_1\,,b_2\,$; it  is obtained adding a dashpot either
 in series to a Voigt model  or in parallel
 to a Maxwell model (Fig. 4c and Fig 4d, respectively).
According to the combination
rule, we add a linear term  to the  Voigt-like creep compliance
and a delta impulsive term to the  Maxwell-like relaxation modulus so that
we obtain $J_e = \infty$ and $G_g = \infty\,. $
We may refer to this model to as the
 {\it anti-Zener} model.
 We have
$$
anti-Zener\;model\;:  \quad
\left[1 +a_1 \, \frac{d}{dt}\right] \sigma(t) =
 \left [  b_1\, \frac{d}{dt} + b_2\, \frac{d^2}{dt^2}\right] \epsilon (t)\,,
\eqno(3.9) $$
and
$$
\!\!
 \cases{
 {\ds J(t) \!= \! J_+t + J_1 \left[ 1-\e^{\ds-t/\tau_\epsilon}\right]},
 & $\!\!
 {\ds J_+ \!=\! \frac{1}{b_1}, \,
 J_1 \!=\!\frac{a_1}{b_1}- \frac{b_2}{b_1^2},\,
       \tau_\epsilon \!=\!\frac{b_2}{b_1}},$ \cr \cr
 {\ds G(t) \!= \! G_-\, \delta (t) + G_1 \,\e^{\ds -t/\tau_\sigma}},
 & $\!\!
 {\ds G_- \!=\! \frac{b_2}{a_1}, \,
   G_1 = \frac{b_1}{a_1}- \frac{b_2}{a_1^2} , \,
       \tau_\sigma = a_1}.$
}
$$
%% \vsp
We point out the condition
$ 0< b_2/b_1< a_1$
in order  $J_1 ,G_1 $ be positive.
 As a consequence,
we note  that, for the {\it anti-Zener} model,
 the relaxation time must be greater than
the retardation time, \ie
$\, 0 < \tau_\epsilon <\tau_\sigma < \infty \,,$
on the contrary of the Zener ($S.L.S.$) model.
 \vsp
%%%%%%%%%%%%%%%%%
In Fig. 4 we exhibit the mechanical representations of the Zener model (3.8),
see a), b), and of the anti-Zener model (3.9)), see c), d).
%%%%%%%%%%% THE END OF FIG 4
\vsp
Based on the combination rule, we can construct
models whose  material functions    are of the following type
$$\cases{
   {\ds J(t)} = {\ds J_g +\sum_{n} J_n \left[
 1-\e^{\,{\ds -t/\tau_{\epsilon,n}}}\right]
        + J_+\, t }\,,\cr
  {\ds G(t)}  = {\ds G_e +  \sum_{n} G_n \,\e^{\,{\ds - t/\tau_{\sigma,n}}}
        + G_-\, \delta (t)}\,,}
         \eqno(3.10) $$
where all the coefficient are non-negative, and
 interrelated because of the
{\it reciprocity relation} (3.3) in the Laplace domain.
We note that the four types of viscoelasticity  of Table 3.1
are obtained from Eqs. (3.10) by taking into account that
$$ \cases{
  J_e <\infty & $\iff \;J_+ =0\,, \q\;J_e =\infty \iff J_+ \ne 0\,,$ \cr\cr
  G_g <\infty &   $\iff G_- =0\,, \q  G_g =\infty \iff G_- \ne 0\,.$
}\eqno(3.11)
$$
Appealing to the theory of  Laplace transforms, we  write
%% and setting $\beta = \sum_n G_n \,. $
$$
\cases{
 s\Js = {\ds J_g + \sum_n \frac{J_n}{ 1+s\,\tau_{\epsilon ,n}}
    + \frac{J_+ }{s} \,,}  \cr\cr
 s\Gs = {\ds (G_e  + \beta) -
 \sum_n \frac{G_n}{1+s\,\tau_{\sigma,n}}
    + G_- \,s} \,,
}
 \eqno(3.12)$$
where we have put $\beta = \sum_n G_n \,. $
%% in order to get analogous expressions.
%%%%%%%%%%%%%%
\vsp
%%%%%%%%%%
Furthermore, as a consequence of (3.12),
 $\,s\Js$ and $s\Gs\,$ turn out to be {\it rational} functions
 in $\CC$ with simple poles and zeros on the negative real axis and,
possibly,
with a simple pole or with a simple zero at $s=0\,, $ respectively.
%%%%%%%%
\vsp 
In these cases the integral constitutive equations (3.1) can be
written in differential form. Following Bland \cite{Bland_60} with
our notations, we obtain for these models
$$ \left[ 1+ \sum_{k=1}^p \,a_k\,{d^k\over dt ^k}\right] \, \sigma (t) =
\left[ m+ \sum_{k=1}^q \,b_k\,{d^k\over dt ^k}\right] \, \epsilon (t)\,,
   \eqno(3.13)$$
where $q$ and  $p$ are integers with $q=p$ or $q=p+1$
and $m, a_k,b_k$ are non-negative constants,
subjected to proper restrictions in order
 to meet the physical requirements of realizability.
The general Eq. (3.13) is referred to as the
{\it operator equation} of the mechanical models.
\vsp
In the Laplace domain,  we thus get
$$ s\Js  =      \rec{s\Gs}= \frac{P(s)}{Q(s)}\,,
   \q {\rm where} \q
   \cases{
   {\ds P(s) = 1+ {\sum_{k=1}^{p}} \, a_k \,s^k\,,} \cr\cr
   {\ds Q(s) = m+ {\sum_{k=1}^{q}} \, b_k \,s^k\,.}
 }
    \eqno(3.14)$$
with $m \ge 0$ and $q=p$ or $q=p+1\,. $
%% $$ s\Js  =
%% \frac{1+{\ds\sum_{k=1}^{p}}\ a_k\,s^k}{ m+{\ds\sum_{k=1}^{q}}\,b_k\,s^k}
%% =  \rec{s\Gs} \eqno(2.  21)$$
The polynomials at the numerator and denominator
 turn out to be  {\it Hurwitz polynomials} (since they have no zeros
 for $\,Re \, \{s\} >0$) whose zeros
are alternating on the negative real axis ($s\le 0$). The least zero
in absolute magnitude is a zero of $Q(s)$.
%% The ratio of any coefficient in $P(s)$ to any coefficient in $Q(s)$ is positive.
The four types of viscoelasticity then correspond to   whether the least zero
is ($J_+ \ne 0$) or is not ($J_+ =0$) equal to zero and to
whether the greatest zero in absolute magnitude is a zero of $P(s)$
($J_g \ne 0$) or a zero of $Q(s)$ ($J_g =0$).
%%%%%%%%%%
\newpage
%%%%%%%%%
\vsp
In Table 3.2 we summarize the four cases, which are expected to occur in the
{\it operator equation} (3.13),  corresponding
to the four  types of viscoelasticity.
%%%%        TABLE 2.2 %%%%%%%%%%%%
\begin{center}
\begin{tabular}{|c||c|c|c|c|c|c|}
\hline
Type & $Order$ & $m$ & $J_g$ & $G_e$ & $J_+$ & $G_-$  \\
\hline
I   & $q=p$   & $>0$ & $a_p/b_p$ & $m$ & $0$     & $0$     \\
II  & $q=p$   & $=0$ & $a_p/b_p$ & $0$ & $1/b_1$ & $0$     \\
III & $q=p+1$ & $>0$ & $  0    $ & $m$ & $0$     & $b_q/a_p$\\
IV  & $q=p+1$ & $=0$ & $  0    $ & $0$ & $1/b_1$ & $b_q/a_p$\\
\hline
\end{tabular}
\vskip 0.25truecm
Table 3.2: The four cases  of the operator equation.
\end{center}
%%%%%%%%%%%%%%%%%%
%% We  can summarize by stating that the general operator equation (2.25)
%% is charcterized by three relevant parameters:
%% $m$ (a non-negative constant) and
%% $p,q$ (non-negative integer numbers with $q=p$ or $q=p+1$).
We recognize that for $p=1$, Eq. (3.13) includes the operator
equations for  the classical models with two parameters: Voigt and
Maxwell, illustrated in Fig. 3, and with three parameters: Zener and
anti-Zener, illustrated in Fig. 4. In fact we recover the Voigt
model (type III )for  $m>0$ and $p=0, q=1$, the Maxwell model (type
II) for $m=0$ and $p=q=1$, the Zener model (type I) for $m>0$ and
$p=q=1$, and the anti-Zener model (type IV) for $m=0$ and $p=1,
q=2$.

%%%%%%%%%%%%%%%%%%%%%%%%% REMARK ON INTIAL CONDITIONS %%%%%%%%%%%%
\vsp
{\sc Remark}.
   We note that the initial conditions at $t=0^+$,
$\sigma^{(h)}(0^+)$ with $h=0,1,\dots p-1$ and $\epsilon^{(k)}(0^+)$
with $k=0,1,\dots q-1$, do not appear in the operator equation but
they are required to be compatible  with the integral equations
(3.1). In fact, since Eqs (3.1) do not contain the initial
conditions,
 some  compatibility conditions at $t=0^+$
must be   {\it implicitly} required both for stress and strain. In
other words, the equivalence between the integral Eqs. (3.1) and the
differential operator Eq. (3.13) implies that when we apply the
Laplace transform to both sides of Eq. (3.13) the contributions from
the initial conditions are vanishing or  cancel in pair-balance.
This can be easily checked for the simplest classical models
described by Eqs (3.6)-(3.9). It turns out that the Laplace
transform of the corresponding constitutive equations does not
contain  any  initial conditions: they are all hidden being zero or
balanced between the RHS and LHS of the transformed equation. As
simple examples let us consider the Voigt model for which $p=0$,
$q=1$ and $m>0$, see Eq. (3.6), and the Maxwell model for which
$p=q=1$ and $m=0$, see Eq. (3.7).
\vsp
For the Voigt model we get %%% in the Laplace domain
$s\bar\sigma(s)= m\bar\epsilon(s) + b \left]s\bar\epsilon(s) -\epsilon(0^+)\right]$,
so,  for any causal stress and strain histories,
it would be
 $$ s\bar J(s) = \frac{1}{m +bs}   \iff \epsilon(0^+)=0\,.\eqno(3.15)$$
We note that the condition $\epsilon(0^+)=0$ is surely satisfied
for any reasonable stress history since $J_g=0$, but
is not valid for any reasonable strain history;
in fact, if we consider the relaxation test for which
$\epsilon(t)=\Theta(t)$ we have $\epsilon(0^+)=1$.
This fact may be  understood recalling that for the Voigt model
we have $J_g=0$ and $G_g=\infty$ (due to the delta contribution in the relaxation modulus).
\vsp
For the Maxwell model we get  %% in the Laplace domain
$\bar\sigma(s)+ a  \left[s\bar\sigma(s) -\sigma(0^+)\right]=
 b\left[s\bar\epsilon(s) -\epsilon(0^+)\right]$,
so,   for any causal stress and strain histories it would be
 $$ s\bar J(s) = \frac{a}{b}+ \frac{1}{bs} \iff a\sigma(0^+)= b \epsilon(0^+)\,. \eqno(3.16)$$
We now note that the condition $a\sigma(0^+)= b \epsilon(0^+)$ is
surely satisfied for any causal history both in stress and in strain.
This fact may be  understood recalling that for the Maxwell model
we have $ J_g >0$ and  $G_g = 1/J_g >0$.
\vsp
Then we can  generalize the above considerations stating that the compatibility
relations of the initial conditions are valid for all the four types of viscoelasticity,
as far as the creep representation is considered.
When the relaxation representation  is considered,
caution is required for the types III and IV, for which, for correctness, we would use the
generalized theory of integral transforms suitable just
for dealing with generalized functions.

%%%%%%%%%%%%%%%%
% \subsection{The time spectral functions} %%% 3.3.
% \vspace*{-8pt} 
\subsection{The time spectral functions}
% \vspace*{-6pt}
  %%%%%%%%%%

From the previous analysis of the classical mechanical models in
terms of a finite number of basic elements, one is led to consider
two {\it discrete} distributions of characteristic times (the {\it
retardation} and the {\it relaxation} times), as stated in (3.10).
However, in more general cases, it is natural to presume the
presence of {\it continuous} distributions, so that, for a
viscoelastic body, the material functions turn out to be of the
following form:
$$
\cases{
   J(t) =  J_g + \alpha \, \int_{0}^{\infty}
  R_\epsilon (\tau)\, \l( 1-\e^{\ds-t/\tau}\r)\, d\tau
        + J_+\, t \,,   \cr\cr
   G(t)  =  G_e +  \beta \, \int_{0}^{\infty}
  R_\sigma  (\tau)\, \e^{\ds-t/\tau}\, d\tau
        + G_-\, \delta (t)\,.
}
\eqno(3.17) $$
where all the coefficients and functions are non-negative. The
function   $  R_\epsilon(\tau) $ is
referred to as the {\it retardation spectrum} while
$  R_\sigma (\tau) $   as the  {\it relaxation  spectrum}.
For the sake of convenience
we shall replace  the suffix $\epsilon$ or $\tau$
with $*$ to denote
anyone of the two   spectra that
we refer simply to as the {\it time-spectral function}.
We require  $R_*(\tau )$  to be  locally integrable
in $\RR^+\,, $ with the supplementary normalization condition
  $ \int_{0}^{\infty}  R (\tau)\, d\tau  = 1$
if the  integral in $\RR^+$ turns out to be  convergent.
\vsp
The discrete distributions of the classical mechanical models,
see (3.10), can be easily recovered from (3.17); in fact,
assuming $\alpha \ne 0\,, \,\beta \ne 0\,, $ we have to put
$$
\cases{
 R_\epsilon (\tau) =
 {\ds \rec{\alpha }\, \sum_n J_n \, \delta (\tau-\tau_{\epsilon,n})\,,
   \q \alpha = \sum_n J_n} \,,  \cr \cr
  R_\sigma (\tau) =
 {\ds \rec{\beta}\, \sum_n G_n \, \delta(\tau-\tau_{\sigma,n})\,,
     \q \beta= \sum_n G_n} \,.
}
   \eqno(3.18)$$
%%%%%%
We devote particular attention to the
time-dependent contributions  to the material functions (3.17)
which are provided by the continuous spectra, \ie
$$
\cases{
 \Psi(t) := {\ds  \alpha\,
   \int_{0}^{\infty}
  R_\epsilon (\tau)\, \l( 1-\e^{\ds-t/\tau}\r)\, d\tau} \,,\cr\cr
  \Phi(t) := {\ds \beta \,
  \int_{0}^{\infty}
  R_\sigma (\tau)\, \e^{\ds-t/\tau}\, d\tau}   \,.
}
   \eqno(3.19)$$
We recognize that
$\Psi(t)$ (that we refer as  the {\it creep function with spectrum})
is a non-decreasing, non-negative function in $\RR^+$
with limiting values $\Psi(0^+)=0 $, $\Psi(+\infty)= \alpha$ or $\infty$,
whereas
$\Phi(t)$ (that we refer as  the {\it relaxation function with spectrum})
is a non-increasing, non-negative function in $\RR^+$
with limiting values $\Phi(0^+)=\beta $ or $\infty$, $\Phi(+\infty)= 0$.
More precisely, in view of their spectral representations (3.19), we have
$$
\cases{
   {\ds \Psi(t) \ge 0\,,\q  (-1)^n \, \frac{d^n\Psi}{dt^n} \le 0}\,, \cr
   {\ds \Phi(t) \ge 0\,,\q  (-1)^n \, \frac{d^n\Phi}{ dt^n}\ge 0}\,,
}
\quad
 t \ge 0 \,,    \q n =1,2, \dots\,.\eqno(3.20)$$
  In other words, $\Phi(t)$ is a {\it completely monotonic} function
  and $\Psi(t)$ is a {\it Bernstein function}
  (namely a non-negative function
with a completely monotonic derivative).
  These properties have been investigated by several authors,
  including Molinari \cite{Molinari 75} and
 more recently by Hanyga \cite{Hanyga_05c}.
%% \vsp %%%%%%%%
The determination of the {\it time-spectral functions} starting
from the knowledge of the creep and relaxation functions is a
problem which can be formally solved through the Titchmarsh inversion
formula of the Laplace transform theory,
see \eg \cite{Gross_53},\cite{Caputo-Mainardi RNC71}.
%% \vsp

 %%%%%%%%%%%%%
% \subsection{Fractional viscoelastic models}
% \vspace*{-8pt} 
\subsection{Fractional viscoelastic models}
% \vspace*{-6pt}
 %%%%%%%%%%%%%%

The straightforward way to introduce fractional derivatives in
linear viscoelasticity is to replace the first derivative in the
constitutive equation (3.5) of the Newton model with a fractional
derivative of order $\nu \in (0,1)$, that, being $\epsilon(0^+)=0$,
may be intended both in the Riemann-Liouville or Caputo sense. Some
people call the fractional model of the Newtonian dashpot with the
suggestive name {\it pot}: we prefer to refer such model to as {\it
Scott-Blair\/} model, to give honour to the scientist who already in
the middle of the past century proposed such a constitutive equation
to explain a material property that is intermediate between the
elastic modulus (Hooke solid) and the coefficient of viscosity
(Newton fluid), see \eg \cite{Scott-Blair 44,Scott-Blair-Caffyn
49,Scott-Blair-Veinoglou-Caffyn 47}. We note that Scott-Blair was
surely a pioneer of the fractional calculus even if he did not
provide a mathematical theory accepted by mathematicians of his
time! 
\vsp
    The use of {\it fractional calculus} in linear viscoelasticity
leads to generalizations of the classical mechanical models:
%% and consequently their operator equation,
 the basic   Newton element  is
substituted by the more general Scott-Blair element
(of order $\nu$).
In fact,   we can construct  the class of these generalized models
from Hooke and Scott-Blair elements,
disposed singly and in branches of two (in series or in parallel).
%%%%%%%%%%%%
Then, extending the procedures
of the classical mechanical models (based on springs and dashpots), we will get the
{\it fractional operator equation}
(that is an operator equation with   fractional derivatives)
 in the form which
properly generalizes (3.13), \ie
$$
\left[1+\sum_{k=1}^p\,a_k\,{d^{\,\nu_k}\over dt^{\,\nu_k}}\right] \,\sigma (t) =
\left[m+\sum_{k=1}^q\,b_k\,{d^{\,\nu_k}\over dt^{\,\nu_k}}\right] \,\epsilon (t)
 \,, \q \nu _k = k + \nu -1\,.   \eqno(3.21)$$
 so, as a generalization of (3.10),
$$ \cases{
  J(t) \,= &
  ${\ds J_g+\sum_{n} J_n\left\{1-\E_\nu \left[-(t/\tau_{\epsilon,n})^\nu\right]\right\}
    + J_+\,{\;t^{\nu} \over \Gamma(1+\nu)}}  \,, $\cr\cr
  G(t) \,= &
  ${\ds G_e +\sum_{n} G_n \,\E_\nu\left[-(t/\tau_{\sigma,n})^\nu\right]
    + G_-\,{\;t^{-\nu} \over \Gamma(1-\nu)}}\,,$
   }
   \eqno(3.22) $$
where all the coefficient are non-negative.
%% \vsp
Of course, for the fractional operator equation (3.21)  the four cases
summarized in Table 3.2 are expected to occur in analogy
with the operator equation (3.13).%%  with the integer-order derivatives
%%%%%%%% FIGURE 5 %%%%%%%%%%%%
\vsp
\begin{figure}[h!]
\begin{center}
 \includegraphics[width=.82\textwidth]{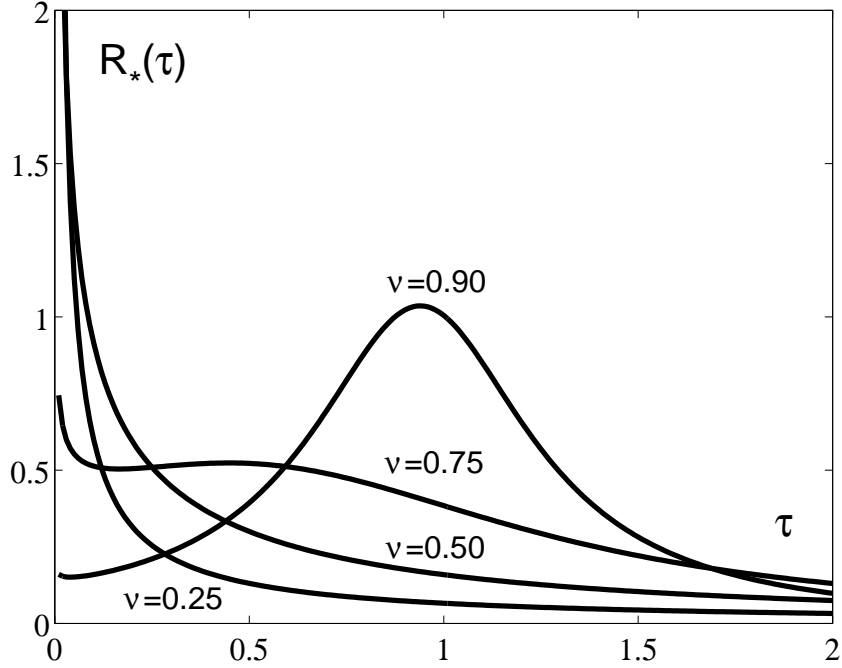}
\end{center}
\caption{Spectral function $R_*(\tau )$
  for  $\nu =0.25,\,0.50,\,0.75,\,0.90$ and  $\tau_*=1$.}
\end{figure}
%%%%%%%%%%  END OF FIG. 5 %%%
\vsp We conclude with some considerations on the presence of the
Mittag-Leffler function in the material functions of the fractional
models. For this purpose let us consider the following {\it creep}
and {\it relaxation} functions with the corresponding {\it
time-spectral} functions
$$
\cases{
\Psi(t) \,= &
$ {\ds \alpha\,\left\{1-\E_\nu\left[-(t/\tau_\epsilon)^\nu \right]\right\}
     =  \alpha \,
   \int_{0}^{\infty} \!\!\!
  R_\epsilon (\tau)\, \left( 1-\e^{\ds-t/\tau}\right)\, d\tau} \,, $\cr\cr
\Phi(t) \,= &
${\ds \beta \,\E_\nu \left[ -(t/\tau_\sigma)^\nu \right]
    = \beta \,     \int_{0}^{\infty} \!\!\!
    R_\sigma(\tau)\, \e^{\ds-t/\tau}\, d\tau} \,. $\cr
  }
   \eqno(3.23)$$
Following Caputo and Mainardi \cite{Caputo-Mainardi PAGEOPH71,Caputo-Mainardi RNC71} and denoting with star
the suffixes $\epsilon \,, \,\sigma \,, $ we obtain
$$
R_*(\tau) =
    \rec{\pi\,\tau  }\,
    {\sin\, \nu \pi \over (\tau /\tau _*)^\nu +
   (\tau /\tau _*)^{-\nu} + 2\, \cos \, \nu \pi } \,.
 \eqno (3.24)$$
From Fig 5 reporting the plots of $R_*(\tau)$  for some values of $\nu$
  we can easily recognize the effect
of the variation of $\nu$  on the character of the spectrum.
For $0<\nu  <\nu _0$, where $\nu  _0 \approx 0.736$
is the solution   of the  equation
$ \nu  = \sin \, \nu  \pi$,
the spectrum $R_*(\tau )$ is a decreasing function
of $\tau $;
subsequently, with increasing $\nu $,   it first exhibits a minimum
and  then a maximum;
for $\nu \to 1 $ it becomes steeper  and steeper near
its maximum  approaching a delta function.
In fact we have
$ {\ds \lim_{\nu \to 1} R_*(\tau)  = \delta (\tau -\tau _*)}$,
where we have set $\tau_*=1$,
being $\tau_*$
the single retardation/relaxation time exhibited by the corresponding
creep/relaxation function.
 Formula (3.24)  was formerly obtained
by Gross \cite{Gross_47} in 1947, when, in the attempt to eliminate
the faults which a power law shows for the creep function, he
proposed the Mittag-Leffler function as an
 empirical law for both the creep and relaxation functions.
In their 1971 papers, Caputo and Mainardi   derived the same result
by introducing into the stress-strain relations a Caputo derivative
that  implies   a {\it memory mechanism} by means of  a convolution
between the first-order derivative and a  power of time, see
(1.6$'$).
%%%%%%%
\vsp
In fractional viscoelasticity governed by the operator equation (3.21)
the corresponding  material functions are obtained
by using the combination rule valid for the classical mechanical models.
Their determination
is made easy if we take into account the following
{\it correspondence  principle} between the classical and fractional
mechanical models, as introduced by Caputo \& Mainardi in 1971.
Taking  $ 0<\nu \le 1$
such correspondence principle can be formally stated by the
following three equations where transitions from Laplace transform pairs
are outlined:
$$
\cases{
{\ds \delta (t) \,\div\, 1
\;\Rightarrow \;
\frac{t^{-\nu}} { \Gamma (1-\nu)}\,\div \, s^{1-\nu}}\,,\cr
%%%%%%%%%%%%%%%%%%%%%%%%%%%%%%
 {\ds  t    \,\div\, \frac{1}{s^2}
 \;\Rightarrow \;
 \frac{t^\nu} {\Gamma (1+\nu )}
   \div \frac{1}{s^{\nu+1}}} \cr
%%%%%%%%%%%%%%%%%%%%%%%%
 {\ds \e^{\ds -t/\tau} \,\div\, \frac{1}{s +1/\tau}
  \;\Rightarrow \;
  \E_\nu [-(t/\tau)^\nu] \div \frac{s^{\nu-1}}{s^\nu +1/\tau}}\,,\cr
}    \eqno(3.25)$$
where $\tau>0$ and
$\,\E_\nu\,$ denotes the Mittag-Leffler function of order $\nu$.
%% We have already referred the reader to  Appendix A for details on this function.

%%%%%%%%%%%%
\vsp
{\sc Remark}.
We note that the initial conditions at $t=0^+$
for the stress and strain
% $\sigma^{(h)}(0^+)$ with $h=0,1,\dots p-1$ and $\epsilon^{(k)}(0^+)$ with
% $k=0,1,\dots q-1$ do not appear in the operator equation
% but they are required to be compatible  with the integral equations (3.1).
do not explicitly enter into the fractional operator equation (3.21)
if they are taken in the same way as  for the classical
mechanical models reviewed in Subsection 3.2. This means
that the approach with  the Caputo derivative, which requires
in the Laplace domain the same initial conditions as the classical
models is quite correct. However, if we assume the same initial conditions,
the approach with the Riemann-Liouville derivative  provides
the same results since, in view of the corresponding Laplace transform rule (1.15),
 the initial conditions do not appear in the Laplace domain.
 The equivalence of the two approaches  has also been noted
 for the fractional Zener model  in a recent note
 by Bagley \cite{Bagley FCAA07}.
   We refer the reader to the paper by Heymans and Podubny \cite{Heymans-Podlubny_06}
   for the physical interpretation of initial conditions
   for fractional differential equations with Riemann-Liouville derivatives,
   especially in viscoelasticity.
   In such field, however, we prefer to adopt the Caputo derivative
   since it requires the same initial conditions
   as in the classical cases, as pointed out in \cite{Mainardi FDA04}:
   by the way,  for  a physical view point,
   these initial conditions  are more   accessible than those required
   in the more general Riemann-Liouville approach, see (1.14).

%%%%%%%%%%%%%% Section 4 %%%%%%%%%%%%%%%%%%%%%%%%%%%%%%%%%%%%%%%%%%%%%%%%%

\section{Some historical notes}

 In this section we provide some
historical notes on how the  Caputo derivative came out as a
contrast to the Riemann-Liouville derivative  and  a sketch about
the generic use of fractional calculus in viscoelasticity.

%%%%%%%%%%%%%
%\subsection{The origins of the Caputo derivative} %% 4.1.
% \vspace*{-8pt}
 \subsection{The origins of the Caputo derivative}
% \vspace*{-6pt}
%%%%%%%%%%%%%%%%%%%%%

 Here we find it worthwhile and
interesting to say something about the commonly used attributes to
Riemann and Liouville and to Caputo for the two types of fractional
derivatives, that have been discussed. \vsp Usually names are given
to pay honour to some scientists who have provided main
contributions, but not necessarily to those  who have as the first
introduced the corresponding notions. Surely Liouville, \eg
\cite{Liouville_1,Liouville_2} (starting from 1832), and then
Riemann \cite{Riemann_1847} (as a student in 1847) have given
important contributions towards    fractional integration and
differentiation, but these notions had previously a story.
As a matter of fact  it was Abel %%  (Abel 1823,Abel 1826)
who solved   his celebrated integral equations
by a fractional integration of order $1/2$ and $\alpha \in (0,1)$, respectively
in 1823 and 1826, see \cite{Abel 23,Abel 26}.
So, Abel,  using the operators that nowadays are ascribed to Riemann and Liouville,
preceded   these eminent mathematicians  by at least 10 years!
\vsp
We remind that
the Laplace transform rule (1.13)
was practically the starting point of Caputo \cite{Caputo 67,Caputo BOOK69}
in defining his generalized derivative in the late Sixties of the past century,
and ignoring the existence of the classical Riemann-Liouville derivative.
Please keep in mind that the first treatise devoted to the so-called fractional
calculus appeared only in 1974, published by  Oldham \& Spanier \cite{Oldham-Spanier BOOK74}
who were unaware of the alternative form (1.6) of the fractional derivative
and of its property (1.13) with respect to the Laplace transform.
 However the form used by Caputo is found in a paper  by Liouville himself
   as recently noted
by Butzer and Westphal \cite{Butzer-Westphal_00} but Liouville
disregarded this notion because he did not recognize its role. \vsp
As far as we know, up to  the middle of the past century the authors
did not take care of the difference between the two forms
(1.5)-(1.6) and of the possible use of  the alternative form (1.6).
Indeed, in the classical book on Differential and Integral Calculus
by the eminent mathematician R. Courant the two forms of the
fractional derivative
 were considered as equivalent,
 see   \cite{Courant BOOK54}, pp.  339-341.
Only in the late Sixties  it seems that the relevance of the
alternative form was recognized. In fact, in 1968
 Dzherbashyan and Nersesian \cite{Dzherbashian-Nersesian_68}
  used the alternative form for dealing with Cauchy problems
  of differential equations of fractional order.
  In 1967, just one year earlier, Caputo
  \cite{Caputo 67}, see also \cite{Caputo BOOK69}, %%%  (Caputo BOOK69)
  used this form   to generalize the usual rule for the Laplace transform
 of a derivative of integer order and to solve some problems
 in Seismology. Soon later, this derivative was adopted by Caputo and Mainardi
 in the framework of the theory of {\it Linear Viscoelasticity},
 see \cite{Caputo-Mainardi PAGEOPH71,Caputo-Mainardi RNC71}.
 \vsp
  Starting with the Seventies many authors, very often ignoring the
  works by Dzherbashian-Nersesian  and Caputo-Mainardi,
  have re-discovered and used the alternative form, recognizing its major utility
  for solving physical problems with standard (namely integer order) initial conditions.
  Although there appeared several papers by different authors,
  including Caputo
  \cite{Caputo 74,Caputo 76,Caputo 79,Caputo 81,Caputo 85,Caputo 86,Caputo 89,%%
  Caputo 93a,Caputo 93b,Caputo 95,Caputo 96a,Caputo 96b,Caputo 98,Caputo LN99},
  Gorenflo et al.
  \cite{Gorenflo CISM97,GoLoLu 02,GorMai CISM97,Gorenflo-Rutman SOFIA94},
  Kochubei \cite{Kochubei 89,Kochubei 90},
  Mainardi \cite{Mainardi ALLOYS94,Mainardi CSF96,Mainardi CISM97,Mainardi-Bonetti 88},
  where the alternative derivative
  was adopted,
   it was mainly  with  the 1999 book by Podlubny \cite{Podlubny 99}
  that it became popular: indeed, in that book it was named as Caputo derivative.
   \vsp
   The notation adopted here was %% formerly
introduced  in a systematic way by
us in our 1996 CISM lectures \cite{GorMai CISM97},
partly based on the 1991 book on Abel Integral Equations
by Gorenflo \&  Vessella \cite{Gorenflo-Vessella LNM91}.

%%%%%%%%%%%%%%%%%%%%%%
%% \subsection{Fractional calculus in  viscoelasticity in the XX-th century} %% 4.2.
% \vspace*{-8pt}
 \subsection{Fractional calculus in  viscoelasticity in the XX-th century} 
%  \vspace*{-6pt}
%%%%%%%%%%%%%%

 Starting from the
 past century a number of authors have (implicitly or  explicitly) used
the {\it fractional calculus} as an empirical method of describing the
properties of visco\-elastic  materials:
 Gemant \cite{Gemant 36,Gemant 50},
Scott-Blair \cite{Scott-Blair-Veinoglou-Caffyn 47,Scott-Blair-Caffyn 49,Scott-Blair 49},
Gerasimov \cite{Gerasimov 48}
were early contributors.
%% in the first half of this century.
%% Scott-Blair used this approach to model the observations made by
%% Nutting (1921-1946) that the stress relaxation phenomenon could be
%% described by fractional powers of time.
%% Scott--Blair noted that fractional - order time derivatives would
%% simultaneously model the observations of Nutting on stress relaxation
%% and those of Gemant on frequency dependence.
Particular mention is due to the theory of hereditary solid mechanics
developed by
Rabotnov \cite{Rabotnov 48,Rabotnov 69,Rabotnov 80},
see also \cite{Rossikhin-Shitikova FCAA07},
that ({\it implicitly}) requires  fractional  derivatives.
\vsp
In the late Sixties, Caputo \cite{Caputo 66,Caputo 67,Caputo BOOK69} and
then Caputo and Mainardi \cite{Caputo-Mainardi PAGEOPH71,Caputo-Mainardi RNC71}
suggested that derivatives of
fractional order (of Caputo type) could be successfully used
to model the dissipation in seismology
and in metallurgy.
\vsp
From then,
applications  of  fractional calculus
in viscoelsticity %% rheology
were considered by an increasing number of authors. Restricting our
attention to a few significant papers published in the past century,
let us quote the contributions  by
%% Atanackovic \cite{Atanackovic CMT02,Atanackovic AM02,Atanackovic ATHENS04},
Bagley \& Torvik \cite{Bagley PhD79,Bagley-Torvik 83,Torvik-Bagley 84},
Caputo \cite{Caputo 74,Caputo 79,Caputo 85,Caputo 89,Caputo 96b,Caputo LN99},
Friedrich  \& associates \cite{Friedrich 91a,Friedrich 91b,Friedrich 93,%%
Friedrich-Braun 94,Friedrich-Schiessel-Blumen 99},
Graffi \cite{Graffi EUROMECH82},
Gaul, Kempfle \& associates  \cite{Beyer-Kempfle ZAMM95,Gaul-Bohlen-Kempfle MRC85,%
Gaul-Klein-Kempfle MRC89,Gaul-Klein-Kempfle MSSP91},
Heymans \cite{Heymans RheolActa96},
Koeller \cite{Koeller 84,Koeller 86},
Mainardi \cite{Mainardi ALLOYS94,Mainardi CISM97},
Meshkov and  Rossikhin \cite{Meshkov-Rossikhin 70},
Nonnenmacher \& associates \cite{Nonnenmacher 91,Nonnenmacher-Glockle 91,Glockle-Nonnenmacher 91},
%% Pipkin \cite{Pipkin BOOK86},
Pritz \cite{Pritz 96,Pritz 98},
Rossikhin and Shitikova  \cite{Rossikhin-Shitikova AMR97}.
\vsp
Additional references
%% to noteworthy contributions related to fractional calculus  in viscoelasticity
up to nowadays can be found in the huge (even not exhaustive)
bibliography of the forthcoming book by Mainardi \cite{Mainardi
BOOK08}. \vsp We like to recall the formal analogy between the
relaxation phenomena in viscoelastic and dielectric bodies; in this
respect the pioneering works by Cole and Cole \cite{Cole-Cole 41,Cole-Cole 42} 
in 1940's  on dielectrics can be considered as
precursors of the implicit use of fractional calculus in that area,
see \eg \cite{Metzler-Klafter 02}.

%%%%%%%%%%%%%%%%%%%%%%%%%%%%%%%%%%%%%%%%%%%%%%%%%%%%%%%%%

%% \section*{Appendix: The functions of Mittag-Leffler type}
\subsection*{Appendix: The functions of Mittag-Leffler type}
% \vspace*{-15pt}

 \subsection*{A.1. The classical Mittag-Leffler function}

The Mittag-Leffler function $E_{\mu}(z)$
(with $\mu>0$)
is an entire  transcendental function of order $1/\mu $, defined
in the complex plane by the power series
$$ E_{\mu} (z) :=
    \sum_{k=0}^{\infty}\,
   {z^{k}\over\Gamma(\mu\,k+1)}\,,
 \q z \in\CC\,. \eqno(A.1)  $$
 It was introduced and studied by the Swedish mathematician
 Mittag-Leffler at the beginning of the XX-th century
 to provide a noteworthy example of entire function
 that generalizes the exponential  (to which it reduces
 for $\mu=1$).
  \vsp
Details on this function  can be  found \eg in the treatises by
Davis \cite{Davis BOOK36},
Dzherbashyan \cite{Dzherbashian BOOK66},
Erdelyi et al. \cite{Erdelyi HTF55},
Kilbas et al. \cite{Kilbas-et-al BOOK06}, %% Gorenflo and Mainardi, 1997;
Kiryakova \cite{Kiryakova BOOK94},
Podlubny \cite{Podlubny  99},
Samko et al. \cite{SKM 93},
Sansone and Gerretsen \cite{Sansone-Gerretsen 60}.
Concerning earlier applications of the Mittag-Leffler function in physics
 let us quote the contributions
 by K. S. Cole, see \cite{Cole 33}, mentioned in the 1936 book
  by Davis, see \cite{Davis BOOK36}, p. 287, in  connection with nerve conduction,
  and by  Gross, see \cite{Gross_47}, in connection with
  creep and relaxation in viscoelastic media.
  \vsp
%% In particular
We  note that the function $E_{\mu} (-x)$ ($x\ge 0$)
 is a completely monotonic function of $x$ if $0< \mu \le 1$,
 as was formerly conjectured by Feller on probabilistic arguments
 and later in 1948 proved by Pollard \cite{Pollard 48}.
  This property  still holds with respect to
 the variable $t$ if we replace   $x$ by  $\lambda \,t^\mu$ ($t\ge 0$) where
 $\lambda$ is a positive constant. Thus in its dependence on $t$  the function
$E_\mu (- \lambda t^\mu ) $ preserves the   {\it complete  monotonicity}
of the exponential $\exp(- \lambda t)$: indeed,
for $0<\mu<1$  it
 is represented in terms of a real Laplace transform (of a real parameter $r$)
of a non-negative   function (that we refer  to as the spectral function)
$$
E_\mu (- \lambda t^\mu)  =
   \frac{1} {\pi}
   \int _0^\infty \!  {\e^{\,\ds -r t}}\,
   \frac{ \lambda r ^{\mu-1}    \,\sin (\mu \pi) }
    {\lambda^2 + 2 \lambda \, r ^\mu \,\cos(\mu\pi)+ r^{2\mu} }
        \,dr\,.
  \eqno(A.2)$$
  We note that as $\mu \to 1^-$ the spectral function tends
  to the generalized Dirac function $\delta(r-\lambda)$.
  \vsp
  We  point out that the Mittag-Leffler function (A.2)
  starts at $t =0$ as a stretched exponential and
   decreases for  $t \to \infty$  %% no longer exponentially but
like a power  with exponent $-\mu $:
$$  E_\mu (-\lambda t^\mu ) \sim
\cases{
{\ds 1 - \lambda \,\frac{t^\mu}{ \Gamma(1+\mu)} }
\sim  {\ds \exp \left\{- \frac{\lambda t^\mu}{ \Gamma(1+\mu)} \right\} }\,,
 & $t \to 0^+\,,$ \cr\cr
{\ds \frac{t^{-\mu }}{ \lambda \,\Gamma(1-\mu)} }\,,
 & $t \to \infty\,.$ \cr
 }
\eqno(A.3)$$
The  integral representation (A.2) and the asymptotic (A.3) can also be derived from
the Laplace transform pair
$$ \L\{ E_\mu(-\lambda t^\mu);s\} = \frac{s^{\mu -1} }{s^\mu + \lambda}\,.\eqno(A.4)$$
In fact  it it sufficient to apply the Titchmarsh theorem for contour
integration in the complex plane
($s = r \e^{i\pi}$) for deriving (A.2) and
the Tauberian theory  ($s\to \infty$ and $s \to 0$) for deriving (A.3).
\vsp
If $\mu =1/2$
we have for $t\ge 0$:
$$
E_{1/2} (-\lambda \sqrt{t})  =
    \e^{\ds \,\lambda^2 t}\, \hbox{erfc} (\lambda \sqrt{t})
\sim 1/(\lambda \sqrt{\pi \,t})\,,\; t\to \infty \,,
 \eqno(A.5)$$
where $ \, \hbox{erfc}\,$ denotes the {\it complementary error}
function, see \eg \cite{AS 65}.  %% (Abramowitz and Stegun, 1965).

%%%%%%%%%  \newpage %\vspace*{-8pt}
  \subsection*{A.2. The generalized Mittag-Leffler function}
 % \vspace*{-6pt}
%%%%%%%%%%%%%%%

 The Mittag-Leffler function  in two parameters
$E_{\mu,\nu}(z) $ ($ \Re \{\mu\} >0$,  $\, \nu \in \CC$)
is  defined %% in the complex plane
by the power series
$$ E_{\mu, \nu} (z) :=
    \sum_{k=0}^{\infty}\,
   \frac{z^k}{\Gamma(\mu \,k + \nu)}\,,
 \q z \in\CC\,. \eqno(A.6) $$
 It generalizes the classical Mittag-Leffler function
 to which it reduces for $\nu=1$.
 It is an entire  transcendental function of order $1/\Re \{\mu\} $
  on which the reader can inform himself  by again consulting
  the treatises cited for the classical Mittag-Leffler function.
% \cite{Erdelyi HTF,Kilbas-et-al BOOK06,GorMai CISM97,Podlubny 99,SKM 93}.
\vsp
The function $E_{\mu, \nu}(-x)$ ($x\ge 0$)
 is completely monotonic in $x$ if $0< \mu \le 1$
 and $\nu \ge \mu$, see \eg
 \cite{Miller-Samko 97,Miller-Samko 01,Schneider 96}.
%%   (Schneider, 1996; Miller and Samko, 1997, Miller and Samko, 2001).
 Again   this property  still holds with respect to $t$
 if we replace the variable   $x$ by $\lambda \,t^\mu$  where
 $\lambda$ is a positive constant.
 In this case the asymptotic representations
 as $t\to 0^+$ and $t\to +\infty$ read
 $$
\null \!\!\! \!
 E_{\mu,\nu}(-\lambda t^\mu)
  \sim
\cases{
{\ds \frac {1}{\Gamma(\nu)}  - \lambda \, \frac {t^\mu}{\Gamma(\nu+\mu)}},
 & $ t \to 0^+,$ \cr\cr
{\ds \frac{1} {\lambda}\, \frac{t^{-\mu +\nu -1}}{\Gamma(\nu-\mu)} }\,, & $ t \to \infty.$ \cr
}
\eqno(A.7)$$
We point out the Laplace transform pair, see \cite{Podlubny 99},
$$ \L\{t^{\nu-1}\, E_{\mu,\nu} (-\lambda t^\mu);s\} =
\frac{s^{\mu -\nu} }{s^\mu + \lambda}\,,  \eqno(A.8)$$
with $\mu, \nu \in \RR^+$.
By aid of this Laplace transform, with $0< \mu = \nu \le 1$,  we can obtain
the useful identity
 $$ t^{-(1-\mu )}
  E_{\mu ,\mu} \lt(- \lambda\, t^{\mu }\rt)
=  - \frac{1}{\lambda}\, \frac{d}{dt} E_\mu  \lt (-\lambda\, t^{\mu }\rt)\,,
\q 0<\mu \le 1\,.
\eqno(A.9)$$
%% In fact we note from (A.8)
To see this it is sufficient to write
$$ \L\lt\{t^{-(1-\mu )}
  E_{\mu ,\mu} \lt(- \lambda\, t^{\mu }\rt)  \rt \}
= \frac{1}{s^\mu + \lambda}= -\frac{1}{\lambda}\,
\lt[ s \frac{s^{\mu-1}}{s^\mu +\lambda} -1  \rt]\,,\eqno(A.10)$$
and invert the Laplace transforms.
Of course the identity (A.9) can be proved directly by differentiating term by term
the power series of the classical Mittag-Leffler function, but, as often  in matters
of fractional calculus, it is simpler to work with the Laplace transform technique.

%%%%%%%%%% THE END OF APPENDIX ON MITTAG-LEFFLER FUNCTIONs

%%%%%%%% \newpage
%%%%%%%%%%%%%%%%%%%%%%%%%%%%%%%%%%%%%%%%%%%%%%%%%%%%%%%%%%%%5


\begin{thebibliography}{99}


\rm


\bibitem{Abel 23}
 Abel, N.H.,
Opl\"osning af et Par Opgaver ved Hjelp af bestemie Integraler
   [Norwegian],
   {\em Magazin for Naturvidenskaberne\/},
   Aargang 1, Bind 2   (1823), 11-27.
  %%  [Christiania 1823]?.
[French translation
%% "Solution de quelques probl\`emes \`a l'aide d'int\'egrales   d\'efinie"
   in: L. Sylov and S. Lie (Editors),
   {\em Oeuvres Compl\`etes de Niels Henrik Abel\/},
  Vol I,   pp. 11-18.
   Christiania  (1881)]
   %%%% from Gorenflo with notes, 8 Sept 00

\bibitem{Abel 26}
  Abel, N. H.,
   Aufloesung einer mechanischen Aufgabe,
  {\it Journal f\"ur die  reine und angewandte Mathematik\/} (Crelle),
  Vol. I (1826), pp. 153-157.
[French translation in: L. Sylov and S. Lie (Editors),
   {\em Oeuvres Compl\`etes de Niels Henrik Abel\/},
  Vol I,   pp. 97-101.
   Christiania  (1881)]

%%   \bibitem[Abel 1826b]{Abel 26b}
 %%  Abel, N. H. (1826b)
 %%  Solution of a mechanical problem [Translated from the
 %%  German with comments by  J.D. Tamarkin]
 %% in D. E. Smith  (Editor),
 %%  {\it A Source Book in Mathematics}
 %% Dover, New York ()1959),  pp.   656-662.  [X]  %%% Gorenflo, Sept 00


 \bibitem{AS 65}
  Abramowitz, M.  and  Stegun, I.A.,
  {\it  Handbook of Mathematical Functions},
 Dover, New York (1965). %% 1965.

\bibitem{Bagley PhD79}
  Bagley,  R.L.,
   {\it Applications of Generalized Derivatives to Viscoelasticity},
  Ph. D. Dissertation, Air Force Institute of  Technology (1979).

 \bibitem{Bagley FCAA07}
  Bagley,  R.L.,
  On the equivalence of the Riemann-Liouville
and the Caputo fractional order derivatives in modeling of linear viscoelastic materials,
{\it Fractional Calculus and Applied Analysis} {\bf 10}, No 2 (2007), 123-126.  %% No 2

\bibitem{Bagley-Torvik 83}
 Bagley, R.L. and  Torvik,  P.J.,
 A theoretical basis for the application  of fractional calculus,
{\it J. Rheology} {\bf 27} (1983), 201-210.

\bibitem{Bagley-Torvik 86}
Bagley, R.L. and  Torvik,  P.J.,
On the fractional calculus model of viscoelastic behavior,
{\it  J. Rheology} {\bf 30} (1986), 133-155.

\bibitem{Beyer-Kempfle ZAMM95}
Beyer, H. and Kempfle, S.,
Definition of physically consistent damping laws with fractional derivatives,
{\it Zeitschrift f\"ur angewandte Mathematk und Mechanik (ZAMM)}
 {\bf 75} (1995), 623-635.

\bibitem{Bland_60}
 Bland, D.R.,
{\it The Theory of Linear Viscoelasticity},
Pergamon, Oxford (1960).

\bibitem{Butzer-Westphal_00}
 Butzer, P.  and  Westphal, U.,
Introduction to fractional calculus,
  In:  Hilfer, H.  (Editor),
{\it Fractional Calculus, Applications in Physics},
 World Scientific, Singapore (2000), pp. 1-85.


\bibitem{Caputo 66}
Caputo, M.,
   Linear models of dissipation whose Q is almost frequency
 independent,   {\it Annali di Geo\-fisica} {\bf 19} (1966), 383-393.

\bibitem{Caputo 67}
 Caputo, M.,
  {Linear models of dissipation whose $Q$ is almost frequency
  independent,  Part II},
  {\it Geophys. J. R. Astr. Soc.} {\bf 13} (1967), 529-539. %%  (1967)


\bibitem{Caputo BOOK69}
  Caputo, M.,
{\it Elasticit\`a e Dissipazione},
  Zanichelli, Bologna (1969). (in Italian)


  \bibitem{Caputo 74}
Caputo, M.,
  Vibrations of an infinite viscoelastic layer with a dissipative memory,
  {\it J. Acoust. Soc. Am.} {\bf 56} (1974), 897-904.

\bibitem{Caputo 76}
Caputo, M.,
Vibrations of an infinite plate with a frequency independent $Q$,
{\it J. Acoust. Soc. Am.} {\bf 60} (1976), 634-639.

\bibitem{Caputo 79}
Caputo, M.,
  A model for the fatigue in elastic materials with frequency
  independent $Q$,
{\it  J. Acoust. Soc. Am.} {\bf 66} (1979), 176-179.

\bibitem{Caputo 81}
 Caputo, M.,
  Elastic radiation from a source in a medium with an almost frequency
  independent $Q$,
{\it  J. Phys.  Earth} {\bf 29} (1981), 487-497.

\bibitem{Caputo 85}
Caputo, M.,
  Generalized rheology and geophysical consequences,
  {\it Tectonophysics} {\bf 116} (1985), 163-172.

\bibitem{Caputo 86}
Caputo, M.,
  Linear and non linear inverse rheologies of rocks,
{\it Tectonophysics} {\bf 122} (1986), 53-71.

\bibitem{Caputo 89}
Caputo, M.,
  The rheology of an anelastic medium studied by means of the
  observation of the splitting of its eigenfrequencies,
{\it  J. Acoust. Soc. Am.} {\bf 86} (1989), 1984-1989.


\bibitem{Caputo 93a}
Caputo, M.,
  The Riemann sheets solutions of anelasticity,
  {\it Ann. Matematica Pura  Appl.} (Ser. IV)
 {\bf 146} (1993), 335-342.

\bibitem{Caputo 93b}
Caputo, M.,
   The splitting of the seismic rays due to dispersion in the Earth's
  interior,
  {\it Rend. Fis. Acc. Lincei} (Ser. IX) {\bf 4} (1993), 279-286.

\bibitem{Caputo 95}
  Caputo, M.,
  Mean fractional-order derivatives differential equations and filters,
  {\it Ann. Univ. Ferrara, Sez VII, Sc. Mat.} {\bf 41} (1995), 73-84.

\bibitem{Caputo 96a}
  Caputo, M.,
  The Green function of the diffusion of fluids in porous media
  with memory,
  {\it Rend. Fis. Acc. Lincei} (Ser. 9) {\bf 7} (1996), 243-250.

\bibitem{Caputo 96b}
  Caputo, M.,
  Modern rheology and dielectric induction: multivalued index of
  refraction, splitting of eigenvalues and fatigue,
  {\it Annali di Geofisica} {\bf 39} (1996), 941-966.

\bibitem{Caputo 98}
Caputo, M.,
  3-dimensional physically consistent diffusion in anisotropic media with
  memory,
  {\it Rend. Mat. Acc.  Lincei} (Ser. 9)  {\bf 9} (1998),  131-14.

 \bibitem{Caputo LN99}
Caputo, M.,
  {\it Lectures on Seismology and Rheological Tectonics},
Lecture Notes, Universit\`a ``La Sapienza", Dipartimento di Fisica,
 Roma (1999).  %%  pp. ???
%% [Down-loadable from
%% []{\tt http://www.uniroma.???}]
%% [Enlarged and revised edition based on the first edition, 1992]

  \bibitem{Caputo-Mainardi PAGEOPH71}
  Caputo, M. and   Mainardi, F.,
A new dissipation model based on memory mechanism, {\it Pure and
Applied Geophysics (Pageoph)} {\bf 91} (1971),  134-147. [Reprinted
in: {\it Fractional Calculus and Applied Analysis} {\bf 10}, No 3
(this issue), 309-324].

\bibitem{Caputo-Mainardi RNC71}
Caputo, M. and  Mainardi, F.,
  Linear models of dissipation in  anelastic solids,
  {\it Rivista del  Nuovo Cimento\/} (Ser. II) {\bf 1} (1971), 161-198.


\bibitem{Carcione-et-al 02}
Carcione, J. M., Cavallini, F.,  Mainardi, F. and Hanyga, A. (2002).
Time-domain seismic modelling of constant-$Q$ wave propagation using
fractional derivatives,
{\it Pure and Appl. Geophys (PAGEOPH)} {\bf 159} (2002), 1719-1736. %%  (2002).
%% This special issue, edited by I. Psenc\'ik and V. Cerven\'y,
%%  contains contributions presented at the international workshop
%%  "Seismic Waves in Laterally Inhomogeneous Media V", Castle of
%%  Zahr\'adki, Czech Republic, June 5-9, 2000. It appears
%% also as an independent book, Birkh\"auser 2002, pp. 511
%%   (Soft-cover, Euro 58, ISBN 3-7643-6677-X)
%%%%%%%%%%%%%%%
 \bibitem{Cole 33}
 Cole,   K.S.,
 Electrical conductance of biological systems,
 Electrical excitation in nerves, In:
{\em Proceedings Symposium on  Quantitative Biology\/}, Cold Spring
Harbor,  New York (1933),  Vol. 1, pp. 107-116.

 \bibitem{Cole-Cole 41}
 Cole, K.S. and  Cole, R.H.,
 Dispersion and absorption in dielectrics,
I. Alternating current characteristics,
{\it J. Chemical Physics} {\bf 9} (1941), 341-349.

\bibitem{Cole-Cole 42}
 Cole, K.S. and  Cole, R.H.,
 Dispersion and absorption in dielectrics,
II. Direct current characteristics,
{\it J. Chemical Physics} {\bf 10} (1942), 98-185.

\bibitem{Courant BOOK54}
Courant, R.,
{\it Differential and Integral Calculus}, Vol. 2,
 London-Glasgow (1954). %%  pp. 339-341,  %%% 1954.

 \bibitem{Davis BOOK36}
 Davis, H.T.,
{\it The Theory of Linear Operators},
  The Principia Press, Bloomington, Indiana (1936).

  \bibitem{Doetsch_74}
  Doetsch, G.,
  {\it Introduction to the Theory and Application of the Laplace
  Transformation}, Springer Verlag, Berlin (1974).

\bibitem{Dzherbashian BOOK66}
 Dzherbashyan, M.M.,
 {\it Integral Transforms and
  Representations of Functions in the Complex Plane}, Nauka, Moscow (1966).
 [in Russian]. There is also the transliteration of the author's name as Djrbashian.

\bibitem{Dzherbashian-Nersesian_68}    %%% Dzherbashian,
 Dzherbashian, M.M. and   Nersesian, A.B.,
Fractional derivatives and the Cauchy problem for differential
 equations of fractional order,
{\it Izv. Acad. Nauk Armjanskvy SSR, Matematika} {\bf 3} (1968),
3-29. (in Russian)

\bibitem{Erdelyi HTF55}
 Erd\'elyi, A., Magnus, W., Oberhettinger,  F and Tricomi, F.G.,
{\it Higher Transcendental Functions} (Bateman Project),
 McGraw-Hill, New York (1955),
 Vol.3, Ch. 18: Miscellanea Functions, pp. 206-227.



    %%%%%%%%%%% Friedrich %%%%%%%%%
    \bibitem{Friedrich 91a}
  Friedrich, Chr.,
Relaxation functions of rheological constitutive equations with
  fractional derivatives: thermodynamic constraints,
in:  Casaz-Vazques, J. and  Jou, D. (Editors),
{\it Rheological Modelling: Thermodynamical and Statistical Approaches},
Springer Verlag, Berlin (1991), pp.  321-330.
    [Lectures Notes in Physics, Vol. 381]
%%%%%%% [Escola de Thermodinamica de Bellaterra, Spain, 24-28 Sept. 1990.]

\bibitem{Friedrich 91b}
  Friedrich, Chr.,
Relaxation and retardation functions of the Maxwell model
  with  fractional derivatives,
 {\it Rheol. Acta} {\bf 30} (1991), 151-158.

%% \bibitem{Friedrich 92}
%%  Friedrich, Chr.,
%%  Rheological material functions for associating comb-shaped or
%%  H-shaped polymers. A fractional calculus approach,
%%   {\it Phil. Mag. Lett.} {\bf 66} (1992), 287-292.    %% (1992)

\bibitem{Friedrich 93}
  Friedrich, Chr.,
Mechanical stress relaxation in polymers: fractional integral
  model versus  fractional differential model,
  {\it J.  Non-Newtonian Fluid Mech.} {\bf 46} (1993), 307-314.  %%5 (1993)



%% `\bibitem{Friedrich-Braun 92}
%%   Friedrich, Chr. and Braun, H. (1992).
%% Generalized Cole-Cole behavior and its rheological relevance,
%% {\it  Rheol. Acta} {\bf 31} No 4 (1992), 309-322.  %%  (1992)

\bibitem{Friedrich-Braun 94}
  Friedrich, Chr. and Braun, H.,
Linear viscoelastic behaviour of complex materials: a fractional mode
representation,
{\it Colloid Polym Sci.}  {\bf 272} (1994), 1536-1546. %% 1994.

\bibitem{Friedrich-Schiessel-Blumen 99}
  Friedrich, Chr., Schiessel, H. and Blumen, A.,
Constitutive behavior modeling and fractional derivatives, In:
Siginer, D.A., Kee, D. and Chhabra, R.P. (Editors), {\it Advances in
the Flow and Rheology of Non-Newtonian Fluids},
Elsevier, Amsterdam (1999), pp.  429-466.  %%  (1999)



\bibitem{Gaul-Bohlen-Kempfle  MRC85}
Gaul, L., Bohlen, S. and  Kempfle, S.,
Transient and forced oscillations of systems with constant hysteretic damping,
{\it Mechanics Research Communications} {\bf 12} (1985), 187-201.%% No 4 (1985)

% \bibitem{Gaul-Kempfle-Klein  ZAMM90}
% Gaul, L., Kempfle, S. and  Klein, P.,
% Transientes Schwingungsverhalten bei der D\"ampfungsbeschreibung
% mit nicht ganzzahligen Zeitableitungen,
% {\it Z. angew. Math. Mech. (ZAMM)} {\bf 70} (1990), T139-T141.%% No 4 (1990)

\bibitem{Gaul-Klein-Kempfle  MRC89}
Gaul, L., Klein, P. and  Kempfle, S.,
Impulse response function of an oscillator with fractional derivative
in damping description,
{\it Mechanics Research Communications} {\bf 16} (1989), 297-305.%% No 5 (1989)

\bibitem{Gaul-Klein-Kempfle  MSSP91}
Gaul, L., Klein, P. and  Kempfle, S.,
Damping description involving fractional operators,
{\it Mechanical Systems and Signal Processing} {\bf 5} (1991), 81-88.%% No 2 (1991)

\bibitem{Gaul-Schanz  MECC97}
Gaul, L. and  Schanz, M.,
Calculation of transient response of viscoelastic solids based on inverse transformation,
{\it Meccanica} {\bf 32} (1997), 171-178.
%%%%%%%%%%% SPRINGER LINK (ex KLUWER)


\bibitem{Gelfand-Shilov_64}
Gel'fand, I.M. and  Shilov, G.E.,
{\it Generalized Functions}, Vol. 1,
 Academic Press, New York (1964).

\bibitem{Gemant 36}
  Gemant, A.,
A method of analyzying experimental results obtained from elastiviscous bodies,
 {\it Physics}  {\bf 7} (1936),  311-317.  %% No 8 (1938)


\bibitem{Gemant 38}
  Gemant, A.,
On fractional differentials,
 {\it Phil. Mag.} (Ser. 7) {\bf 25} (1938),  540-549.

\bibitem{Gemant 50}
 Gemant, A.,
 {\it Frictional Phenomena},
 Chemical Publ. Co, Brooklyn N.Y. (1950).
%% quoted by  Koeller  1986 (Acta Mech: Polynomial Operators
%% Gemant and Scott-Blair were early contributors in the
%% use of fractional calculus to study phenomenological
%% constitutive equations for material behavior !!!!!!!!


\bibitem{Gerasimov 48}
  Gerasimov, A.,
A generalization of linear laws of deformation
and its applications to problems of internal friction,
{\it Prikl. Matem. i Mekh.} {\bf 12} (1948), 251-260. [in Russian]

    \bibitem{Glockle-Nonnenmacher 91}
Gl\"ockle, W. G. and  Nonnenmacher, T. F. (1991).
Fractional integral operators and Fox functions in the theory
 of viscoelasticity,
{\it Macromolecules} {\bf 24} (1991), 6426-6434.


\bibitem{Gorenflo CISM97}
Gorenflo, R.,
  Fractional calculus: some numerical methods,
   In:   Carpinteri, A. and  Mainardi, F. (Editors)
{\it Fractals and Fractional Calculus in Continuum Mechanics},
 Springer Verlag, Wien (1997), pp. 277-290.
  [Reprinted in  {\tt http://www.fracalmo.org}]

%  \item[-]  %% {Glockle-Nonnenmacher 94}
% Gl\"ockle, W. G. and  Nonnenmacher, T. F. (1994).
% Fractional  relaxation and the time-temparature superposition principle,
% {\it Reological Acta}, {\bf 33},  337-343. %%  (1994);

\bibitem{GoLoLu 02}
Gorenflo, R.,  Loutschko, J. and  Luchko, Yu.,
Computation of the  Mittag-Leffler  function $E_{\alpha, \beta} (z)$
and its derivatives,
{\em Fractional Calculus and Applied Analysis\/}
{\bf  5}, No 4 (2002),  491-518.  %% No 4 (2002)


\bibitem{GorMai CISM97}
  Gorenflo, R. and  Mainardi, F.,
 Fractional calculus:
  Integral and differential equations of fractional order,
  In: A. Carpinteri and F. Mai\-nardi (Editors),
  {\em Fractals and Fractional Calculus in Continuum Mechanics\/},
  Springer Verlag, Wien (1997),   pp. 223-276.
  [Reprinted in  {\tt http://www.fracalmo.org}]

   \bibitem{Gorenflo-Rutman SOFIA94}
 Gorenflo, R.  and Rutman, R.,
On ultraslow and intermediate processes, In:
 P. Rusev, I. Dimovski and V. Kiryakova  (Editors),
 {\it Transform Methods and Special Functions, Sofia 1994},
 Science Culture Technology Publ., Singapore (1995), pp. 171-183.
%% [Proc. Workshop on Transform Methods and Special Functions,
%%    Sofia, Bulgaria, 12-17 August 1994]
%%%%%%%%%%%%%%%%%%%%%%%%%%%%%%
\bibitem{Gorenflo-Vessella LNM91}
 Gorenflo, R. and Vessella, S.,
 {\it Abel Integral Equations: Analysis and Applications},
 Springer Verlag, Berlin (1991).  [Lecture Notes in Mathematics No 1461].

\bibitem{Graffi EUROMECH82}
  Graffi, D.,
  Mathematical models and waves in linear viscoelasticity,
  In: F. Mainardi (Editor),
 {\it Wave Propagation in Viscoelastic Media},
 Pitman, London (1982), pp. 1-27.
[Research Notes in Mathematics, Vol. 52]


%%%%%%%%%%%%%%
\bibitem{Gross_47}
 Gross, B.,
On creep and relaxation,
{\it J. Appl. Phys.} {\bf 18} (1947), 212-221.
%%%%%%%%%%%%%%%
\bibitem{Gross_53}
 Gross, B.,
{\it Mathematical Structure of the Theories of Viscoelasticity},
Hermann \& C., Paris (1953).
%%%%%%%%%%%%%%%%%%%%%%%%%%%%%%%%%%%
% \bibitem{Hanyga_05a}
% Hanyga, A.,
% Realizable constitutive equations in linear viscoelasticity,
% in   Le M\'ehaut\'e, A.,   Tenreiro Machado, J.A.,
% Trigeassou, J.C. and  Sabatier, J. (Editors),
% {\it Fractional Derivatives and Their Applications},
% U-BOOKS Verlag, Neus\"ass (2005), pp. 353-364.
% [Selected Papers at the {\it First IFAC Workshop on Fractional Differentiation and its
%     Applications (FDA'04)}, Bordeaux (France) 19-21 July 2004]
%%% BOOK of the Proceedings pp.  68-73.
% \bibitem{Hanyga_05b}
% Hanyga, A.,
% Physically acceptable viscoelastic models,
% in    Hutter, K. and Y. Wang, Y. (Editors),
% {\it Trends in Applications of Mathematics to Mechanics},
% Proceedings {\it  Workshop STAMM'04,   Seheim (Germany) 22-28 August 2004},
% Shaker Verlag GmbH, Aachen (2005), 12 pp.
% [{\tt www.geo.uib.no/hjemmesider/andrzej/index.html}]
%%%%%%%%%%%
\bibitem{Hanyga_05c}
Hanyga, A.,
Viscous dissipation and completely monotonic relaxation moduli,
{\it Rheologica Acta} {\bf 44} (2005), 614-621.
%%%%%%%%%
\bibitem{Heymans RheolActa96}
Heymans, N.,
Hierarchical models for viscoelasticity: dynamic behaviour
in the linear range,
{\it Rheol. Acta} {\bf 35} (1996), 508-519.
%%%%%%%%%%%%%%%%
\bibitem{Heymans-Podlubny_06}
Heymans, N.  and Podlubny, I.,
Physical interpretation of initial conditions for fractional differential equations
with Riemann-Liouville fractional derivatives,
{\it Rheol. Acta} {\bf 45} (2006), 765-771.   %
%%%%%%%%%%%%%%
\bibitem{Hilfer BOOK00}
 Hilfer, R. (Editor),
{\it Fractional Calculus, Applications in Physics}, World
Scientific, Singapore (2000).
\bibitem{Hilfer 00}
 Hilfer, R.,
Fractional time evolutions,
 In:  Hilfer, H.  (Editor),
{\it Fractional Calculus, Applications in Physics},
 World Scientific, Singapore (2000), pp. 87-130.
%%%%%%%%%%%%%%%%
 \bibitem{Kempfle-Schaefer-Beyer NLD02}
 Kempfle, S., Sch\"afer, I. and Beyer, H.,
 Fractional calculus via functional calculus: theory and applications,
 {\it Nonlinear Dynamics} {\bf 29} (2002), 99-127.
%%%%%%%%%%%%
\bibitem{Kilbas-et-al BOOK06}          %  %
 Kilbas, A.A., Srivastava, H.M.  and  Trujillo, J.J.,
 {\it Theory and Applications of Fractional Differential Equations},
 Elsevier, Amsterdam (2006).
 %%%%%%%%%%%%%%%%
 \bibitem{Kiryakova BOOK94}
  Kiryakova, V.,
   {\it Generalized Fractional Calculus and Applications},
       Longman, Harlow
[Pitman Research Notes in Mathematics, Vol. 301] \& J. Wiley, N.
York (1994).
%%%%%%%%%%%%%
\bibitem{Kochubei 89}
Kochubei,  A.N.,
  A Cauchy problem for evolution equations of fractional order,
  {\it Differential Equations}  {\bf  25} (1989),  967-974.
[English translation from the Russian Journal
{\it Differenttsial'nye Uravneniya}]

\bibitem{Kochubei 90}
 Kochubei, A.N.,
  Fractional order diffusion,
  {\it Differential Equations}  {\bf  26} (1990),  485-492.
[English translation from the Russian Journal
{\it Differenttsial'nye Uravneniya}]
%%%%%%%%%%%%%%

\bibitem{Koeller 84}
Koeller, R.C.,
Applications of fractional calculus to the theory of  viscoelasticity,
{\it J. Appl. Mech.} {\bf 51} (1984),299-307.

\bibitem{Koeller 86}
Koeller, R.C.,
Polynomial operators, Stieltjes convolution and fractional calculus
 in hereditary mechanics,
 {\it Acta Mech.} {\bf 58} (1986), 251-264.

 \bibitem{Liouville_1}
 Liouville, J., Memoire sur quelques questions de geometrie et de
 mecanique et sur un nouveau genre de calcul pour resudre ces questions,
 {\it J. Ecole Polytech} {\bf 13} (21) (1832), 1-69.


 \bibitem{Liouville_2}
 Liouville, J., Memoire sur le changemant de variables dans calcul des
 differentielles d'indices quelconques,
 {\it J. Ecole Polytech} {\bf 15} (24) (1835), 17-54.


 \bibitem{Magin BOOK06}
Magin, R.L.,
{\it Fractional Calculus in Bioengineering},
Begell Houuse Publishers, Connecticut (2006).


\bibitem{Mainardi ALLOYS94}
Mainardi,  F.,
Fractional relaxation  in anelastic solids,
{\it Journal of Alloys and Compounds} {\bf 211/212} (1994), 534-538.

\bibitem{Mainardi CSF96}
   Mainardi, F.,
  Fractional relaxation-oscillation and fractional
  diffusion-wave phenomena,
  {\it Chaos, Solitons and Fractals\/} {\bf 7} (1996), 1461--1477.
%%%%%%%%%%%%%%%%%%%%%%%%%%
\bibitem{Mainardi CISM97}
 Mainardi, F.,
   Fractional calculus:
 some basic problems in continuum and statistical mechanics,
In: A. Carpinteri and F. Mainardi (Editors), {\em Fractals and
Fractional Calculus in Continuum Mechanics\/},
 Springer Verlag, Wien and New-York (1997),  pp. 291-348.
 [Reprinted in {\tt http://www.fracalmo.org}]
%%%%%%%%%%%%%%%%%%
\bibitem{Mainardi FDA04}
 Mainardi, F.,
 Physical and mathematical aspects  of fractional calculus
 in linear viscoelasticity, In:
 A. Le M\'ehaut\'e, J.A. Tenreiro Machado, J.C. Trigeassou, J. Sabatier
 (Editors),
  Proceedings the 1-st IFAC Workshop on
{\it Fractional Differentiation and its applications}(FDA'04), pp. 62-67
[ENSEIRB, Bordeaux (France), July 19-21, 2004]
%%%%%%%%%%%%
\bibitem{Mainardi BOOK08}
 Mainardi, F.,
   {\it Fractional Calculus and Waves in Linear Viscoelasticity}
   Imperial College Press, London (2008), to appear.
%%%%%%%%%%%%%
\bibitem{Mainardi-Bonetti 88}
 Mainardi, F. and  Bonetti, E.,
The application of real-order derivatives in linear visco-elasticity,
{\it  Rheologica Acta} {\bf 26} Suppl. (1988), 64-67.


% \bibitem{Mainardi-Gorenflo JCAM00}
% Mainardi, F. and  Gorenflo, R.,
% On Mittag-Leffler-type functions in fractional evolution processes,
% {\it  J. Comput.  Appl. Math.} {\bf 118} (2000),   283-299.

%%%%%%%%%%%%%%%%%5
\bibitem{Meshkov-Rossikhin 70}
Meshkov, S.I. and Rossikhin, Yu.A., 
Sound wave propagation in a viscoelastic medium whose hereditary properties are determined by
weakly singular kernels, 
In: Rabotnov, Yu.N. (Editor), 
{\it Waves in Inelastic Media}, Kishniev (1970), pp. 162-172. [in Russian]

%%%%%%%%%%%%%%%%%

\bibitem{Metzler-Klafter PhysRep00}
Metzler, R.  and  Klafter, J., The random walk's guide to anomalous
diffusion: a fractional dynamics approach, {\it Physics Reports}
{\bf 339} (2000), 1-77.

\bibitem{Metzler-Klafter 02}
Metzler, R. and  Klafter, J.,
 From stretched exponential to inverse power-law:  fractional dynamics,
 Cole-Cole relaxation processes, and beyond,
   {\it J. Non-Crystalline Solids} {\bf 305} (2002),  81-87.
\bibitem{Molinari 75}
 Molinari,  A.,
Visco\'elasticit\'e lin\'eaire et function
compl\`etement monotones,
{\it Journal de M\'ecanique} {\bf 12} (1975),  541-553.

\bibitem{Miller-Ross BOOK93}
 Miller, K.S.  and   Ross, B.,
 {\it An Introduction to the Fractional
  Calculus and Fractional Differential Equations},
Wiley, New York (1993).

\bibitem{Miller-Samko 97}
Miller, K.S. and Samko, S.G.,
A note on the  complete monotonicity of the generalized Mittag-Leffler
function,
 {\it Real Anal. Exchange} {\bf 23} (1997),  753-755.  %% No 2
%%%%%%%%%%%%%%%%%%%%%%%%%%%
\bibitem{Miller-Samko 01}
Miller, K.S. and Samko, S.G.,
Completely monotonic functions,
 {\it Integr. Transf. and Spec. Funct.} {\bf 12} (2001),  389-402.%% No 4
%%%%%%%%%%%%%%%%%%%%%%%%%%%%
\bibitem{Nonnenmacher  91}
  Nonnenmacher, T.F.,
Fractional relaxation equations for viscoelasticity and
    related phenomena,
 In:  Casaz-Vazques, J. and  Jou, D. (Editors),
{\it Rheological Modelling: Thermodynamical and Statistical Approaches},
Springer Verlag, Berlin (1991), pp. 309-320.
    [Lectures Notes in Physics, Vol. 381]


\bibitem{Nonnenmacher-Glockle 91}
Nonnenmacher, T.F.  and  Gl\"ockle,  W.G.,
A fractional model for mechanical stress relaxation,
{\it Phil. Mag. Lett.} {\bf 64} (1991),   89-93.  %% No 2 (1991)


\bibitem{Nonnenmacher-Metzler 95}
  Nonnenmacher, T.F.  and  Metzler, R.,
 On the Riemann-Liouville  fractional calculus and some
recent applications,
{\it Fractals} {\bf 3} (1995), 557-566.

\bibitem{Oldham-Spanier BOOK74}
  Oldham, K.B.  and J. Spanier, J.,
 {\it The Fractional Calculus},
   Academic Press, New  York (1974).  %%  (1974).

\bibitem{Pipkin BOOK86}
 Pipkin, A.C.,
{\it Lectures on Viscoelastic Theory},
Springer Verlag, New York (1986),  2-nd Edition. [1-st ed. 1972]
%% (Applied Mathematical Sciences No 7)


\bibitem{Podlubny 99}
   Podlubny, I.,
  {\it Fractional Differential Equations},
  Academic Press, San Diego (1999).

\bibitem{Podlubny MATLAB06}
   Podlubny, I.,
    {\it Mittag-Leffler function}.
   MATLAB Central, File exchange \\(2006),
  [{\tt http://www.mathworks.com/matlabcentral/fileexchange}]
  %% /loadFile.do?objectId=8738&objectType=FILE]
  %%%%%%%%%%
\bibitem{Pollard 48}
        Pollard, H.,
    The completely monotonic character of the Mittag-Leffler function
     $E_\alpha (-x)\,,$
   {\em Bull. Amer. Math. Soc.\/} {\bf 54} (1948), 1115--1116.
%%%%%%%%%%
%%%%%%%%% PRITZ %%%%%%%%%
\bibitem{Pritz 96}
Pritz, T.,
Analysis of four-parameter fractional derivative model of real solid materials,
{\it J. Sound and Vibration} {\bf 195} (1996), 103-115. %%% No 1, 8 Aug 1996

%% \bibitem[Pritz (1996b)]{Pritz 96b}
%% Pritz, T. (1996b).
%% Dynamic Young's modulus and loss factor of floor covering materials,
%% {\it Applied Acoustics} {\bf 49}, 179-190. %%% No 2, October 1996.

\bibitem{Pritz 98}
Pritz, T.,
Frequency dependences of complex moduli and complex Poisson's ratio
 of real solid materials,
{\it J. Sound and Vibration} {\bf 214} (1998),  83-104. %%% No 1, 2 Jul 1998

\bibitem{Pritz 99}
Pritz, T.,
Verification of local Kramers-Kronig relations
for complex modulus by means of
fractional derivative model,
{\it J. Sound and Vibration} {\bf 228} (1999), 1145-1165. %%% No 5, 16 Dec 1999

%% \bibitem[Pritz (2000)]{Pritz 00}
%% Pritz, T. (2000).
%% Measurement methods of complex Poisson's ratio of viscoelastic materials,
%% {\it Applied Acoustics} {\bf 60}, 279-292. %%% No 3, July 2000.

% \bibitem[Pritz (2001)]{Pritz 01}
% Pritz, T. (2001).
% Loss factor peak of viscoelastic  materials: magnitude to width relations,
% {\it J. Sound and Vibration} {\bf 246}, 265-280. %%% No 2, 13 Sept 2001.

% \bibitem[Pritz (2003)]{Pritz 03}
% Pritz, T. (2003).
% Five-parameter fractional derivative model for polymeric damping  materials,
% {\it J. Sound and Vibration} {\bf 265}, 935-952. %%% No 5, 28 Aug 2003.


\bibitem{Rabotnov 48}
 Rabotnov,   Yu. N.,
Equilibrium of an elastic medium with after effect,
{\it Prikl. Matem. i Mekh.} {\bf 12} (1948), 81-91. [in Russian]

 \bibitem{Rabotnov 69}
 Rabotnov,   Yu. N.,
{\it Creep Problems in Structural Members},
 North-Holland, Amsterdam (1969).
 [English translation of  the 1966 Russian edition]

\bibitem{Rabotnov 80}
 Rabotnov,   Yu. N.,
 {\it Elements of Hereditary Solid Mechanics},
  MIR, Moscow (1980).
 [English translation, revised from the 1977 Russian edition]
%%%%%%%%%%%%%%%%%%%%%

 \bibitem{Riemann_1847}
 Riemann, B.,
 Versuch einer allgemeinen Auffassung der Integration und Differentiation,
 {\it Bernard Riemann: Gesammelte Mathematische Werke}, Teubner Verlag, Leipzig (1892),
 XIX, pp. 353-366.
 [Reprinted in {\it Bernard Riemann: Collected Papers}, Springer Verlag, Berlin (1990)
 XIX, pp. 385-398.]

\bibitem{Rossikhin-Shitikova AMR97}
Rossikhin, Yu.A.  and  Shitikova, M.V.,
Applications of fractional calculus to dynamic problems
of linear and nonlinear fractional mechanics of solids,
{\it Appl. Mech. Review} {\bf 50} (1997),  15-67.  %% (1997)  No. 1.
%%%%%%%%%%%
\bibitem{Rossikhin-Shitikova FCAA07}
Rossikhin, Yu.A.  and  Shitikova, M.V.,
Comparative analysis of viscoelastic models involving fractional derivatives
of different orders,
 {\it Fractional Calculus and Applied Analysis} {\bf 10}, No 2 (2007), 111-121.

 \bibitem{SKM 93}
  Samko, S.G.,  Kilbas, A.A. and  Marichev, O.I.,
 {\it Fractional Integrals and Derivatives: Theory  and  Applications},
 Gordon and Breach, New York (1993).
 [Translation from the Russian edition, Nauka i Tekhnika, Minsk (1987)]

 \bibitem{Sansone-Gerretsen 60}
  Sansone, G. and Gerretsen, J.,
{\it Lectures on the Theory of Functions of a Complex Variable},
 Vol. I. {\it Holomorphic Functions},
 Nordhoff, Groningen (1960).  %%  (1960) pp. 488.

%\\ {\footnotesize{[See \S 6.13 Mittag-Leffler function, pp. 345-349,
% \S 8.4 The Mittag-Leffler summability of a power series, pp. 431-438.%
% Gawromski (1988) quotes this book (p. 143 \S 3.7.3:
% Expansions   of Bernoulli's and Euler's numbers in infinite series)
% for the expansion of cotan $x$ in terms of the Riemann $\zeta$ function.
% More precisely the Riemann $\zeta$ function is treated in \S 7.5, pp. 364-367]}}
 %%%%%%%%%%%%%%%%%%%%%%%%%%%%%%%%%%
\bibitem{Schneider 96}
 Schneider, W.R.,
 Completely monotone generalized Mittag-Leffler functions,
 {\it Expositiones Mathematicae} {\bf 14} (1996),  3-16.

  \bibitem{Scott-Blair 44}
 Scott-Blair, G.W.,
 Analytical and integrative aspects  of the stress-strain-time problem,
{\it J. Scientific Instruments}  {\bf 21} (1944), 80-84.

\bibitem{Scott-Blair 49}
  Scott-Blair,  G.W.,
{\it Survey of General and Applied Rheology},
 Pitman, London (1949). %%  1949.
%%  quoted by GROSS  1953 book and by Koeller


\bibitem{Scott-Blair-Caffyn 49}
 Scott-Blair, G.W. and  Caffyn, J.E.,
 An application  of the theory of quasi-properties to the treatment
  of anomalous stress-strain relations,
{\it Phil. Mag.} [Ser. 7] {\bf 40} (1949), 80-94.

\bibitem{Scott-Blair-Veinoglou-Caffyn 47}
Scott-Blair, G.W., Veinoglou, B. C. and  Caffyn, J.E.,
Limitations of the Newtonian time scale in relation to non-equilibrium
rheological states and a theory of quasi-properties,
{\it Proc. R. Soc. London A} {\bf 189} (1947), 69-87.

 \bibitem{Seybold-Hilfer FCAA05}
 Seybold, H.J. and Hilfer, R.,
 Numerical results for the generalized Mittag-Leffler function,
 {\it Fractional Calculus and Applied Analysis} {\bf 8} (2005), 127-139.

\bibitem{Sokolov-Klafter-Blumen FK02}
Sokolov, I,  Klafter, J. and A. Blumen, A.,
Fractional kinetics,
{\it Physics Today} {\bf 55} (2002), 48-54.


\bibitem{Torvik-Bagley 84}
 Torvik, P.J.  and  Bagley, R.L.,
On the appearance of the fractional derivatives in the behavior of
  real materials,
{\it J. Appl. Mech.} {\bf 51} (1984), 294-298.


\bibitem{West BOOK03}
West, B.J., Bologna, M. and  Grigolini, P.,
{\it Physics of Fractal Operators}, Springer Verlag, New York (2003).
%% [Institute for Nonlinear Science]

\bibitem{Zaslavsky BOOK05}
Zaslavsky,  G.M.,
{\it Hamiltonian Chaos and Fractional Dynamics},
Oxford University Press, Oxford (2005).

\bibitem{Zener BOOK48}
Zener, C.,
{\it Elasticity and Anelasticity of Metals},
  University of Chicago Press, Chicago (1948).
%%  \end{itemize}
 \end{thebibliography}
\end{document}